\documentclass[onecollarge,natbib]{svjour3}
\bibpunct{[}{]}{;}{n}{}{,} 
\smartqed  
\usepackage{graphicx}
\usepackage[normalem]{ulem}
\usepackage{amsmath}
\usepackage{amssymb}
\usepackage{mathtools}
\usepackage{color}
\usepackage{xcolor}
\usepackage{lineno}
\usepackage{hyperref}
\usepackage{fullpage}
\usepackage{relsize}
\usepackage{mathrsfs}
\usepackage{bm}
\usepackage{cancel}
\usepackage{soul}
\usepackage{booktabs}
\usepackage{caption}
\usepackage[caption=false]{sub fig} 
\usepackage{multirow}

\colorlet{darkgreen}{green!50!black}
\colorlet{brightyellow}{yellow!75!red}
\colorlet{orange}{red!50!yellow}
\colorlet{darkblue}{blue!60!black}
\colorlet{darkred}{red!80!black}

\newcommand{\ket}[1] {{\left.|#1\right>}}

\journalname{ }
\begin{document}
\title{Hadron Spectra, Decays and Scattering Properties within Basis Light Front Quantization
       \thanks{Presented at LightCone 2017, Mumbai, India.}
}
\titlerunning{BLFQ - Progress and Prospects}        
\author{James~P.~Vary   \and
        Lekha Adhikari        \and
		Guangyao~Chen    \and
		Shaoyang~Jia        \and
		Meijian~Li	              \and 
        Yang~Li                   \and
        Pieter~Maris           \and
        Wenyang~Qian      \and
        John~R.~Spence   \and
        Shuo Tang              \and
        Kirill~Tuchin	     \and
        Anji~Yu       	     \and
        Xingbo~Zhao
}
\authorrunning{Vary et al.} 
\institute{James~P.~Vary \and Lekha Adhikari \and Guangyao~Chen
\and Meijian~Li \and  Pieter~Maris 
\and Wenyang~Qian \and John~R.~Spence  
\and Shuo Tang \and Kirill~Tuchin \and Anji~Yu \and Xingbo~Zhao 
\at Department of Physics and Astronomy, Iowa State University, Ames, IA 50011, USA \\
\email{jvary@iastate.edu} 
\and
		Yang~Li \at 
		Department of Physics, College of William and Mary, Williamsburg, VA 23168, USA
		\and
        Xingbo~Zhao \at
        Institute of Modern Physics, Chinese Academy of Sciences, Lanzhou, 730000, China \\
}
\date{\today}
\maketitle
\begin{abstract}
We survey recent progress in calculating properties of the electron and hadrons within the Basis Light Front Quantization (BLFQ) approach. We include applications to electromagnetic and strong scattering processes in relativistic heavy ion collisions. We present an initial investigation into the glueball states by applying BLFQ with multigluon sectors, introducing future research possibilities on multi-quark and multi-gluon systems.
\end{abstract}
\section{Introduction}\label{introduction}
Static and dynamic properties of the hadrons, including finite nuclei, are the foci of major theoretical and experimental efforts in the nuclear physics community. Interesting topics include the nonperturbative roles of the constituent quarks and gluons in the manifested phenomena such as the distribution of angular momenta, pileup of gluon distributions at low longitudinal momentum fractions, electromagnetic moments/transitions, emergence of exotic structures beyond the simple constituent models and diffractive production cross sections. Our BLFQ framework addresses these phenomena with a relativistic treatment of the Hamiltonian developed with input from the QCD Lagrangian supplemented by confining terms. We solve for the mass eigenstates and their light-front amplitudes in the Basis Light Front Quantization (BLFQ) approach \cite{Vary:2009gt,Honkanen:2010rc,Vary:2011np}. The light front amplitudes bridge our theory with observables such as form factors, decay constants and those in time-dependent scattering processes \cite{Zhao:2013jia,Zhao:2013cma}. Specifically, our Hamiltonian framework allows the study of time-dependent phenomena using the time-dependent Basis Light Front Quantization (tBLFQ) approach. We gradually improve the tBLFQ results by replacing modeled background fields with those obtained directly from QCD. 

In this article we overview recent developments, expanding our report of last year \cite{Vary:2016ccz}, and survey prospects for the near future. For additional perspectives of recent research and future prospects in light-front
Hamiltonian theory, see Refs.~\cite{Bakker2013.165,Hiller:2016itl} and references therein.

\section{Basis Light-Front Quantization\label{sec 2}}

The non-perturbative solution of quantum field theory within the Hamiltonian
framework has a rich history.  Pioneering efforts addressed the mass 
eigenstate problem of the light-front Hamiltonian in the discretized 
plane-wave basis~\cite{Brodsky98.299} called Discretized Light-Cone Quantization (DLCQ), whose application continues to the present~\cite{Hiller:2016itl}.  
In order to better address the conserved total angular momentum projection, to 
improve numerical convergence when confining interactions are present 
and to facilitate multi-fermion and multi-boson applications, we introduced BLFQ by adapting successful 
methods from {\it ab initio} nuclear structure theory \cite{Vary:2009gt}.

We aim to solve the light-front mass eigenvalue problem
expressed as, 
\begin{linenomath*}
\begin{equation}
( \hat P^+ \hat P^- -\hat P_{\perp}^2)|\psi_h\rangle = M^2_h  |\psi_h\rangle, \label{eq:effective_hamiltonian}
\end{equation}
\end{linenomath*}
where the operators $\hat{P}^+$ and $\hat{P}^-$ are the longitudinal momentum ($+$) and the light-front quantized Hamiltonian ($-$), while $\hat{P}_{\perp}$ represents the transverse momentum.
The invariant-mass ($M$) spectrum and light-front state vectors $|\psi_h\rangle$ result from diagonalizing Eq.~\ref{eq:effective_hamiltonian} in a suitable matrix representation of the effective Hamiltonian. Expressing the 
state vectors in either coordinate or momentum space yields the light-front wavefunctions (LFWFs).

The Fock space expansion for $|\psi_h\rangle$ in a chosen basis representation results in a sparse-matrix eigenvalue problem.
By choosing the two-dimensional (2D) 
harmonic-oscillator (HO) for the transverse modes, one benefits from the developments of the no-core shell model (NCSM) used successfully in nuclear many-body theory \cite{Barrett:2013nh,Navratil:2000ww,Navratil:2000gs,Maris:2008ax} while retaining a fully covariant framework. An appealing feature of BLFQ is the ability to factorize transverse center-of-mass motion from the relative coordinates (momenta), resulting in the preservation of transverse boost invariance of the LFWFs~\cite{Vary:2009gt,Li:2013cga,Maris:2013qma}. We note that the choice of the 2D-HO for the transverse basis space is harmonious with the phenomenologically successful light-front AdS/QCD soft-wall Hamiltonian for hadrons \cite{Brodsky09.081601,Brodsky15.1}. 

BLFQ has been applied to QED problems at strong coupling such as the electron anomalous 
magnetic moment \cite{Honkanen:2010rc,Zhao:2014xaa}, non-linear Compton scattering 
\cite{Zhao:2013jia,Zhao:2013cma} and the positronium spectrum \cite{Wiecki:2014ola}.
Here, we briefly summarize recent BLFQ applications to bound-state and scattering problems:
electron form factors~\cite{Tang:2018aaa} in Sect.~\ref{sec 3},
heavy quarkonium structure and radiative transitions \cite{Li:2015zda,Li:2017mlw,Li:2018aaa,Adhikari:2018} in Sect.~\ref{sec 4},
mixed-flavor heavy quarkonium \cite{Tang:2018bbb} in Sect.~\ref{sec 5}, 
light mesons \cite{Qian:2018aaa} in Sect.~\ref{sec 6}, baryon systems \cite{Yu:2018aaa} in Sect.~\ref{sec 7}, 
quark jet scattering in a color glass condensate~\cite{Zhao:2013jia,Zhao:2013cma,Chen:2017uuq,Li:2018bbb} 
in Sect.~\ref{sec 8}, 
and an initial application to the glueball sector in Sect.~\ref{sec 9},
each of which points pathways to future developments and applications.
A promising approach to non-perturbative renormalization in the light front quantization approach is also
presented at this meeting~\cite{Zhao:2014hpa,Zhao_this_meeting}. 
\section{Electron Form Factors in BLFQ} \label{sec 3}

\begin{figure}
\centering 
\begin{minipage}{0.49\textwidth}
    \includegraphics[width=\linewidth]{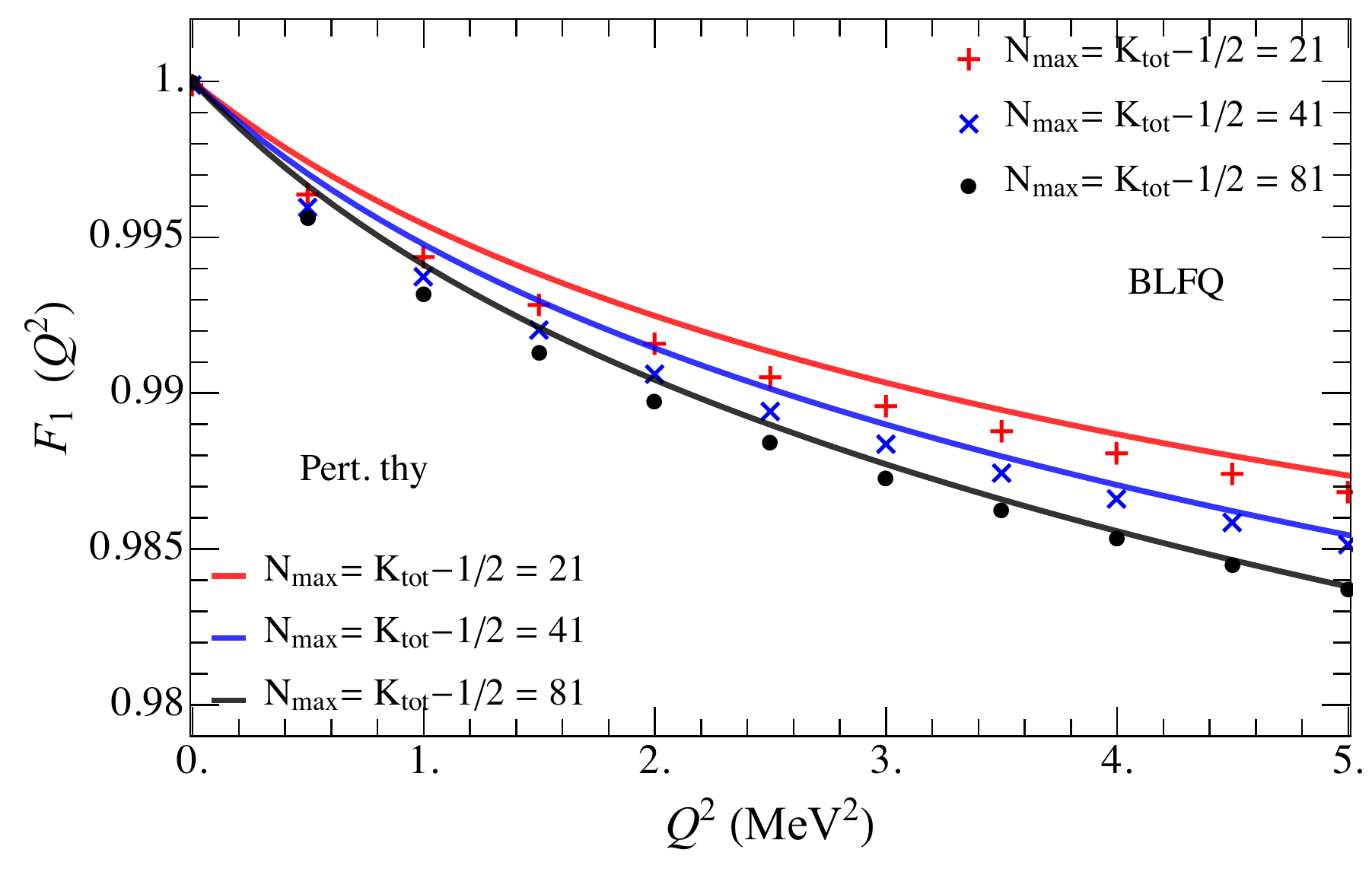}
    \end{minipage}
    \hspace{\fill} 
    \begin{minipage}{0.49\textwidth}
    \includegraphics[width=\linewidth]{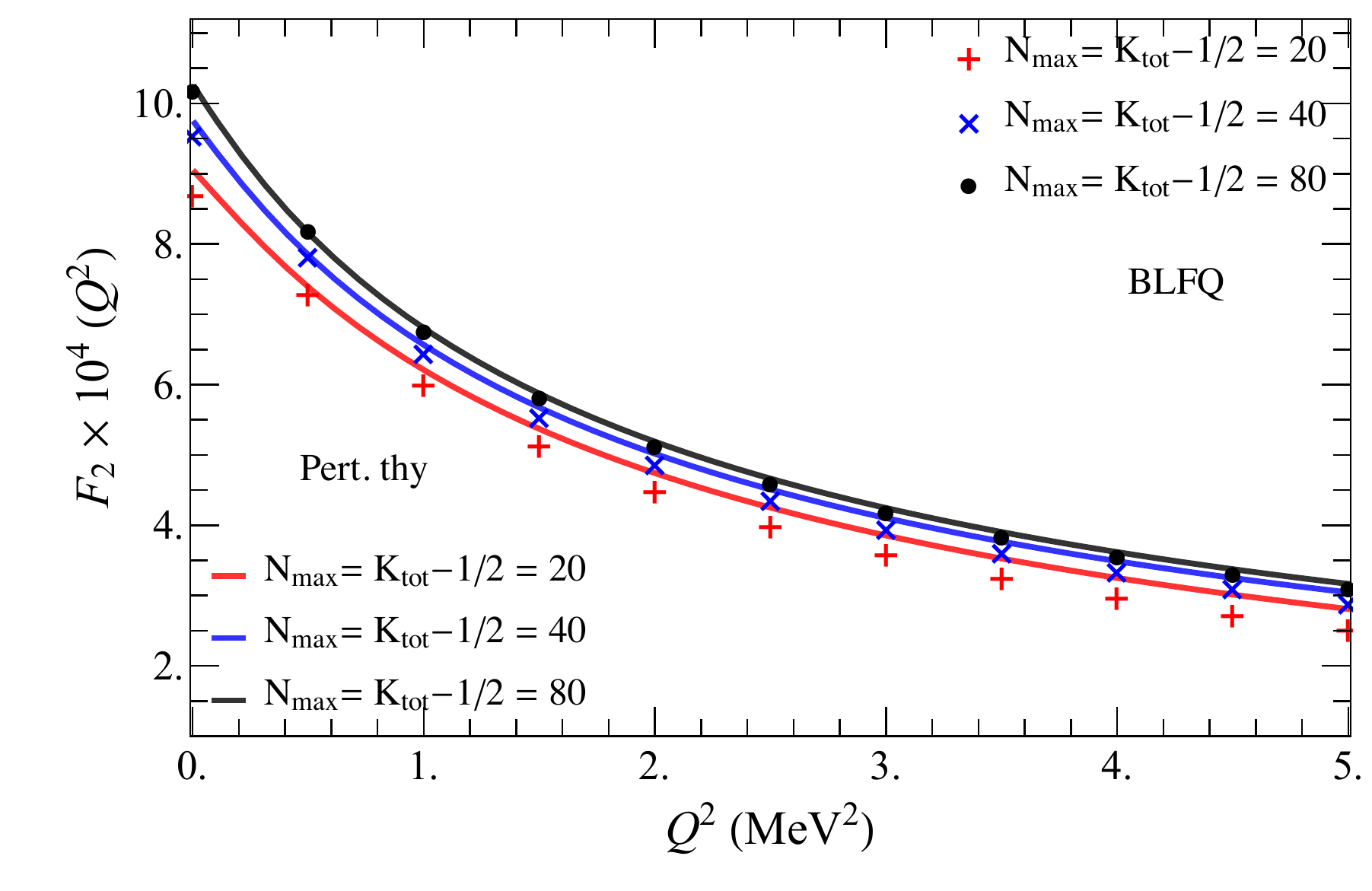}
    \end{minipage}
    \begin{minipage}{0.49\textwidth}
    \includegraphics[width=\linewidth]{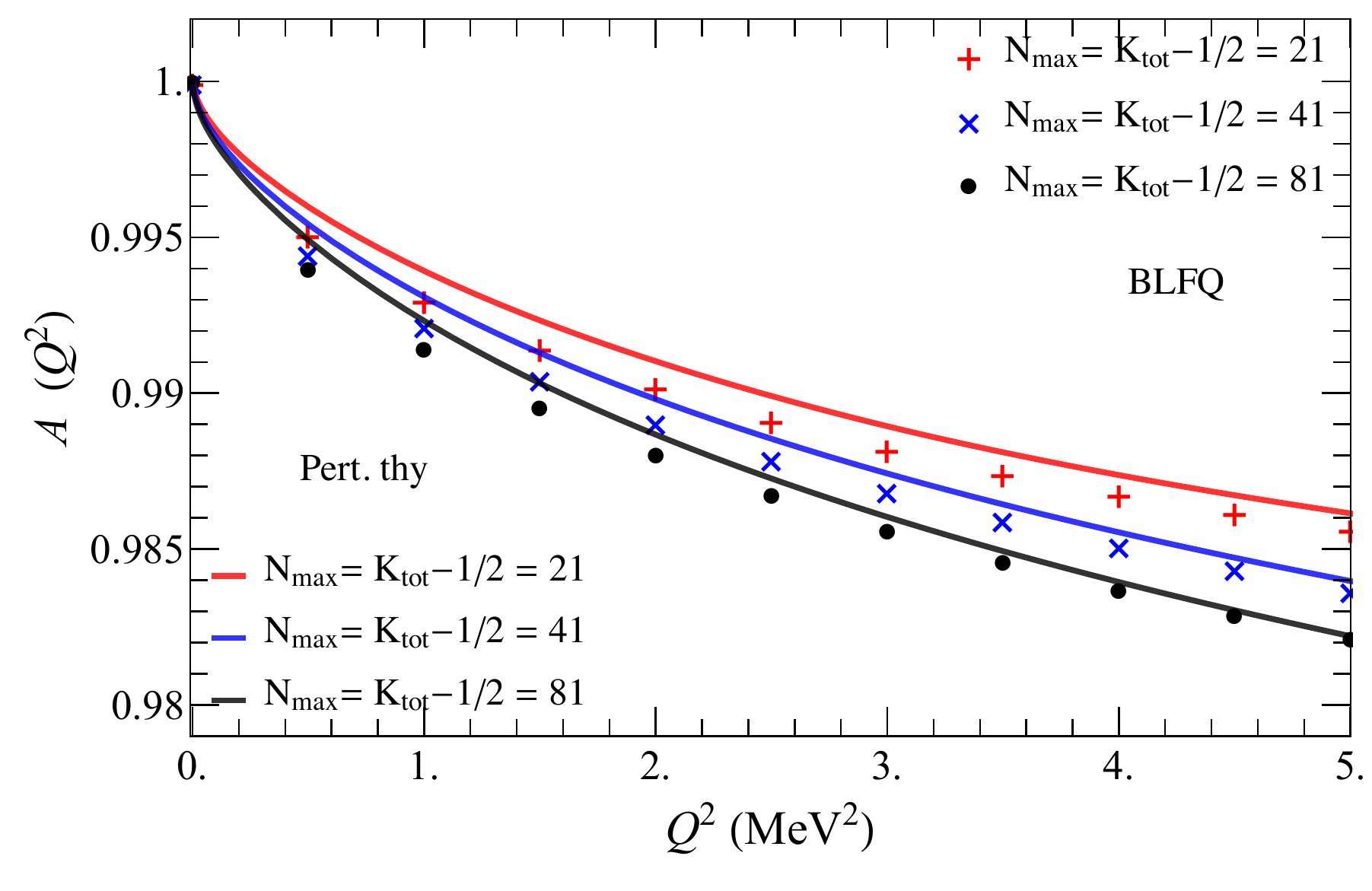}
    \end{minipage}
    \hspace{\fill} 
    \begin{minipage}{0.49\textwidth}
    \includegraphics[width=\linewidth]{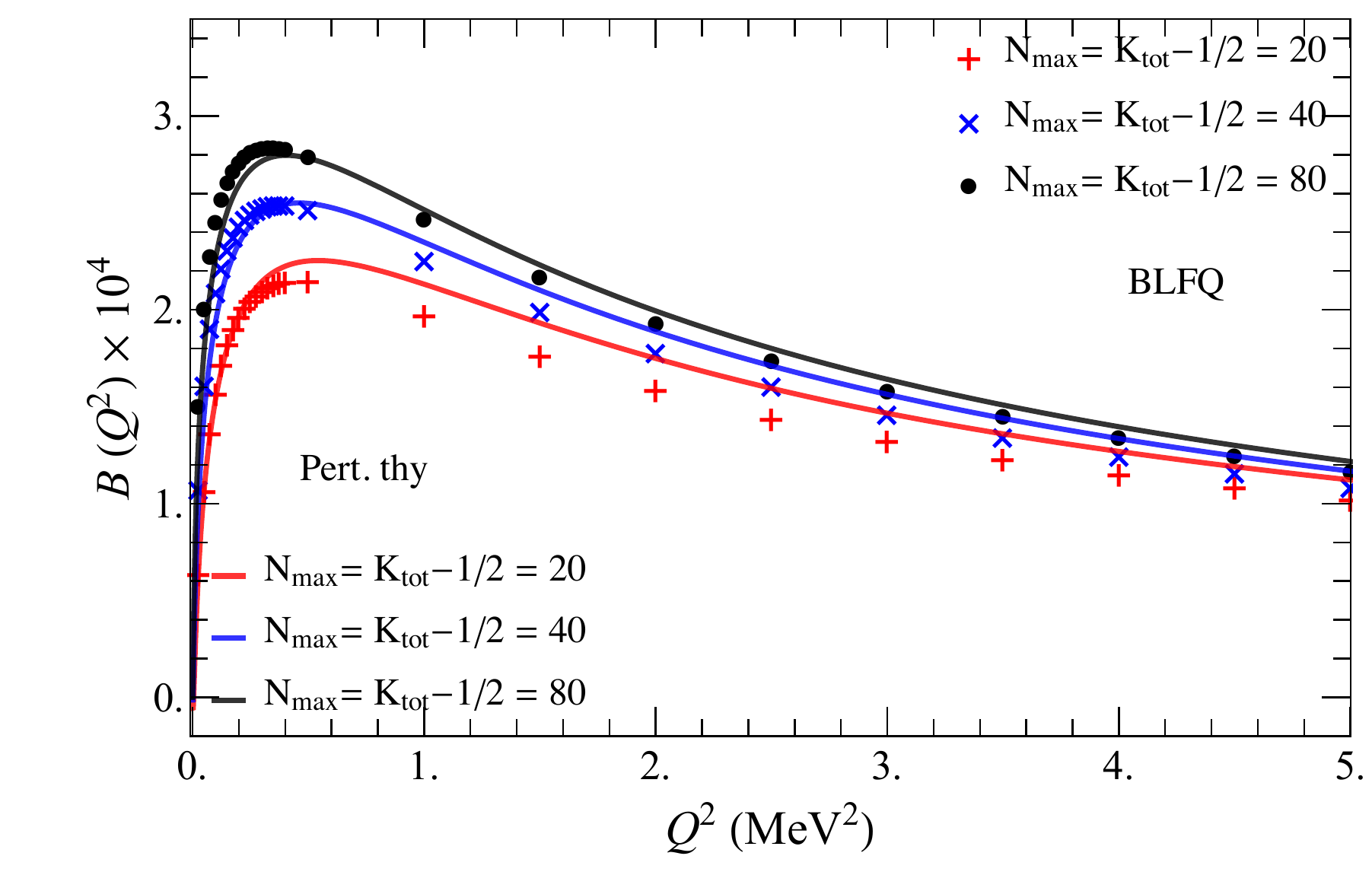}
    \end{minipage}
\caption{(Color online) 
Electron form factors obtained from BLFQ (discrete points) and LFPT (solid lines)
as a function of $Q^2$.  Top panel: Dirac ($F_1$) and Pauli ($F_2$) form factors; 
bottom panel: gravitational form factors $A= A_f + A_b$ and $B = B_f + B_b$.
The BLFQ regulators, $N_{\max}$ and $K_{\mathrm{tot}}$, are given 
in the legends. The corresponding LFPT regulators are defined
in the text.
}
\label{electronff}
\end{figure}

By truncating the QED Fock space into
one electron and one electron plus one photon sectors, we obtain non-trivial electromagnetic
and gravitational form factors for the physical electron~\cite{Tang:2018aaa}.
We present results for these form factors at selected values of the finite basis space cutoff regulators using BLFQ in Fig.~\ref{electronff} (discrete points).  
The regulator $N_{\rm max}$ defines the upper limit of the sum of 2D-HO quanta
for the transverse basis states and the regulator $K_{\mathrm{tot}}$ defines the upper limit of the 
longitudinal plane wave modes, for which we adopt antiperiodic (periodic) boundary 
conditions for fermions (bosons); i.e. $K_{\mathrm{tot}}$ is the sum of the electron's
(half-odd-integer) and the photon's (integer) longitudinal quanta.

For comparison, we present the same quantities calculated with light-front perturbation theory
(LFPT) in Fig.~\ref{electronff} (continuous curves) using cutoff regulators in momentum space that are matched 
to the basis space regulators of BLFQ.  That is, after introducing $b$ as the momentum scale of the 
underlying 2D-HO basis, we define the matched infrared (ultraviolet) transverse cutoff of LFPT as 
$b /\sqrt{N_{\rm max}}$ ($b \sqrt{N_{\rm max}}$). For the infrared (ultraviolet) 
longitudinal momentum cutoff of LFPT, we adopt $m_{\mathrm{e}}/ K_{\mathrm{tot}}$ ($m_{\mathrm{e}}K_{\mathrm{tot}}$) with $m_{\mathrm{e}}$ taken as the physical electron mass.

We present the electron's Dirac form factor 
$F_1(Q^2)$, Pauli form factor $F_2(Q^2)$, as well as the spin-conserving and spin-flip
gravitational form factors $A(Q^2)= A_f(Q^2) + {A_b(Q^2)}$ 
and $B(Q^2) = {B_f(Q^2) + B_b(Q^2)}$ in Fig.~\ref{electronff} through a sequence of basis space regulators. For $F_1$ and $A$ we
averaged over results obtained at $N_{\rm max}$ values adjacent to the value
quoted in the legend
to smooth over numeric staggering that is not found in the other form factors. 
This staggering of the BLFQ results in $F_1$ decreases with higher $N_{\rm max}$
so we adopt this averaging procedure to simplify the comparison with LFPT.

We obtain the expected agreement between LFPT and DLCQ in all results as shown in Fig.~\ref{electronff}.  Furthermore, we demonstrate
a good convergence pattern and improving agreement between LFPT and BLFQ 
as regulators are lifted. All form factors exhibit the expected result as $Q^2 \rightarrow 0$.
That is, $F_1$ approaches unity and $F_2$ approaches the electron's anomalous 
magnetic moment. In addition, the anomalous gravitomagnetic moment vanishes, 
i.e. $B(Q^2=0) = 0$ \cite{Brodsky:2000ii} at all choices of the regulators.

\section{Heavy Quarkonium Structure and Radiative Transitions} \label{sec 4}

Our effective Hamiltonian in the quark-antiquark $(q\bar{q})$ Fock space \cite{Li:2015zda,Li:2017mlw} 
is based, in part, on the massless 2-dimensional holographic QCD 
Hamiltonian \cite{Brodsky09.081601,Brodsky15.1}. We define our effective Hamiltonian as

\begin{linenomath*}
 \begin{equation}
  H_\text{eff} = \frac{\bm k^2_\perp + m_q^2}{x} + \frac{\bm k^2_\perp + m_{\bar{q}}^2}{1-x} + \kappa^4 x(1-x)\bm
r^2_\perp + V_\text{g} + V_L(x)\label{eq:BLFQ_effective_hamiltonian}
 \end{equation}
\end{linenomath*}
where, $x = p^+_q/P^+$ is the longitudinal momentum fraction of the quark, 
$\mathbf{k}_\perp = \mathbf{p}_{q\perp} - x \mathbf{P}_\perp$ is the 
intrinsic transverse momentum, and $\mathbf{r}_\perp = \mathbf{r}_{q\perp} - \mathbf{r}_{\bar{q}\perp}$ 
is the transverse separation of the quark and the anti-quark. 
Our effective Hamiltonian incorporates the masses of the quark $m_q$ and the antiquark $m_{\bar q}$,
longitudinal dynamics with a confining term $V_L$ \cite{Li:2015zda} and 
one-gluon exchange $V_\text{g}$ with either a fixed coupling~\cite{Li:2015zda} or a running coupling~\cite{Li:2017mlw}, as specified within Refs.~\cite{Li:2015zda,Li:2015zda,Li:2017mlw}.
Our longitudinal confining potential is given by
\begin{linenomath*}
\begin{equation}
V_{L}(x) = -\frac{\kappa^4}{(m_q+m_{\bar{q}})^2} \partial_x \big( x (1-x) \partial_x\big) 
\end{equation}
\end{linenomath*}
and differs from other proposals \cite{Glazek11.1933,Trawinski14.074017,Chabysheva13.143}.
Specifically, the resulting longitudinal function for the ground state is very similar to that of the invariant mass ansatz (IMA) of Brodsky and de T\'eramond~\cite{Brodsky15.1}, except for the endpoint behavior. See Ref.~\cite{Li:2017mlw} for a detailed comparison. 
The parameter $\kappa$ fixes both the transverse confining strength in concert with 
holographic QCD \cite{Brodsky09.081601,Brodsky15.1} as well as the strength within $V_L$.
Incorporating the running coupling in $V_\text{g}$ improved the comparison between theoretical and 
experimental spectroscopy while reducing the violation of rotational symmetry~\cite{Li:2017mlw}.

\begin{figure}
\begin{tabular}{cc}
\subfloat[$\chi_{\mathrm{c}0}: 1^3P_0 (0^{++})$]{\includegraphics[scale=0.33]{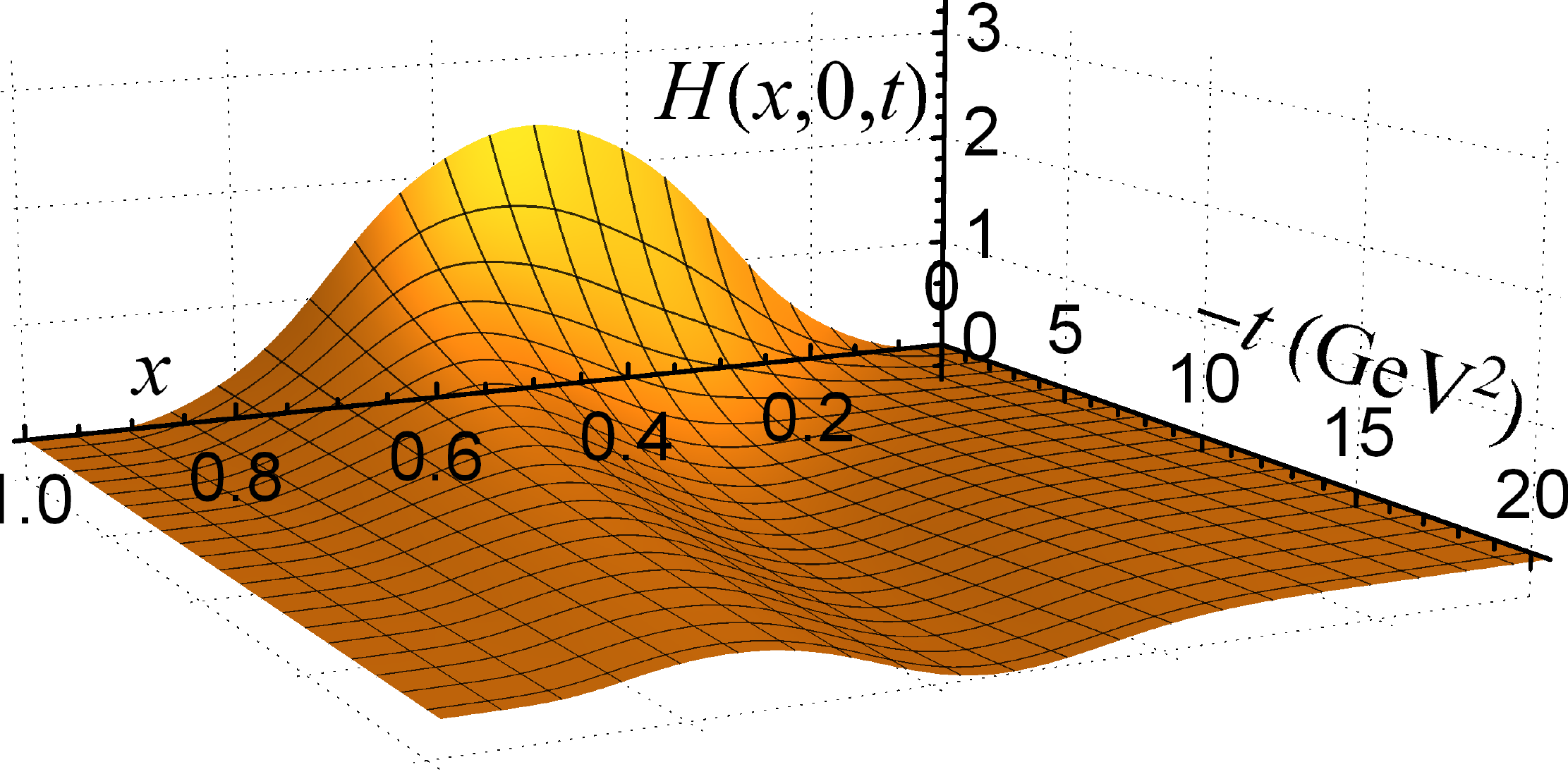}}\\ 
\subfloat[$\chi_{\mathrm{c}0}: 1^3P_0 (0^{++})$]{\includegraphics[scale=0.33]{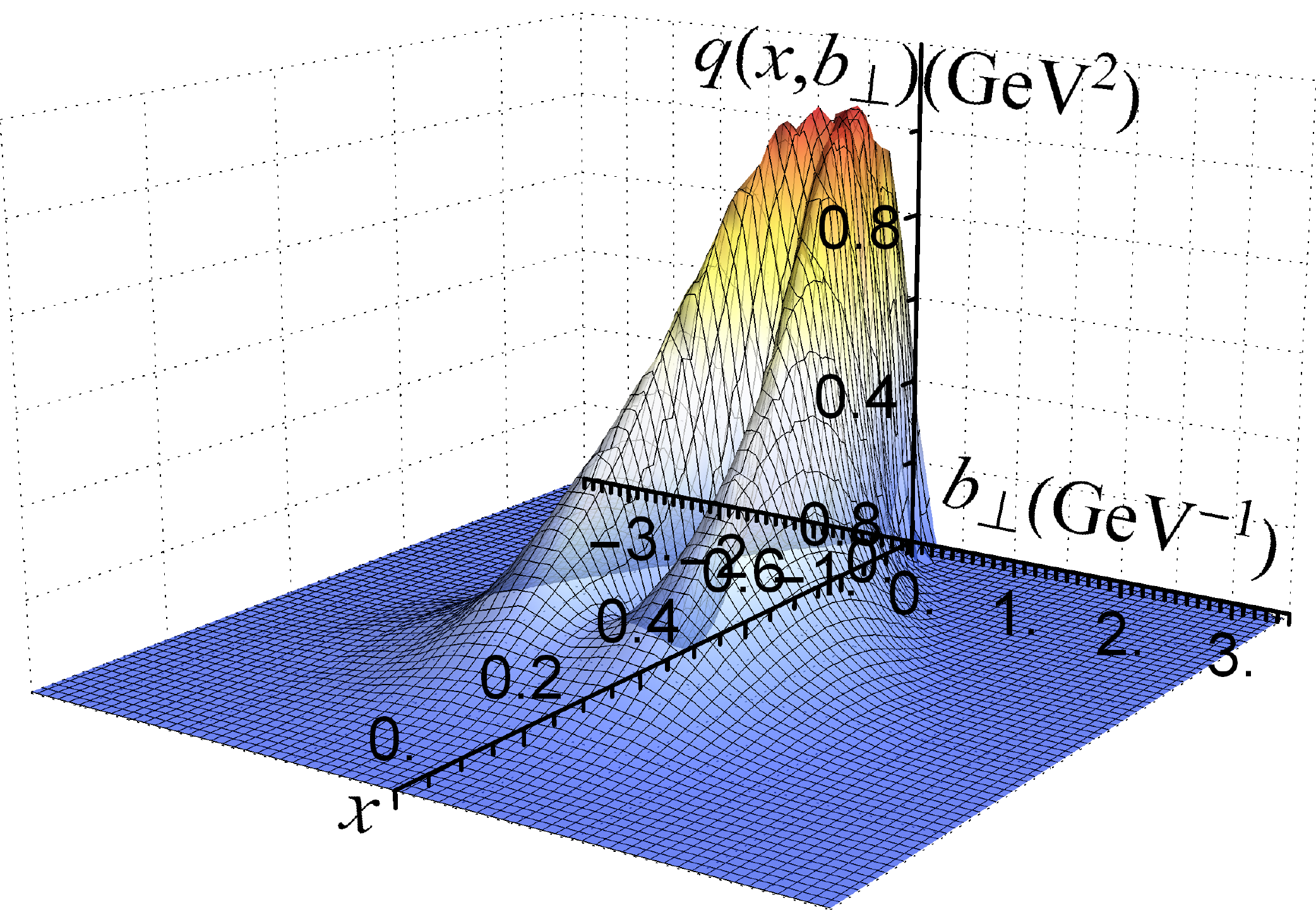}} \\ 
\end{tabular}
\begin{tabular}{cc}
\subfloat[$\eta_{\mathrm{c}}^\prime: 2^1S_0 (0^{-+})$]{\includegraphics[scale=0.33]{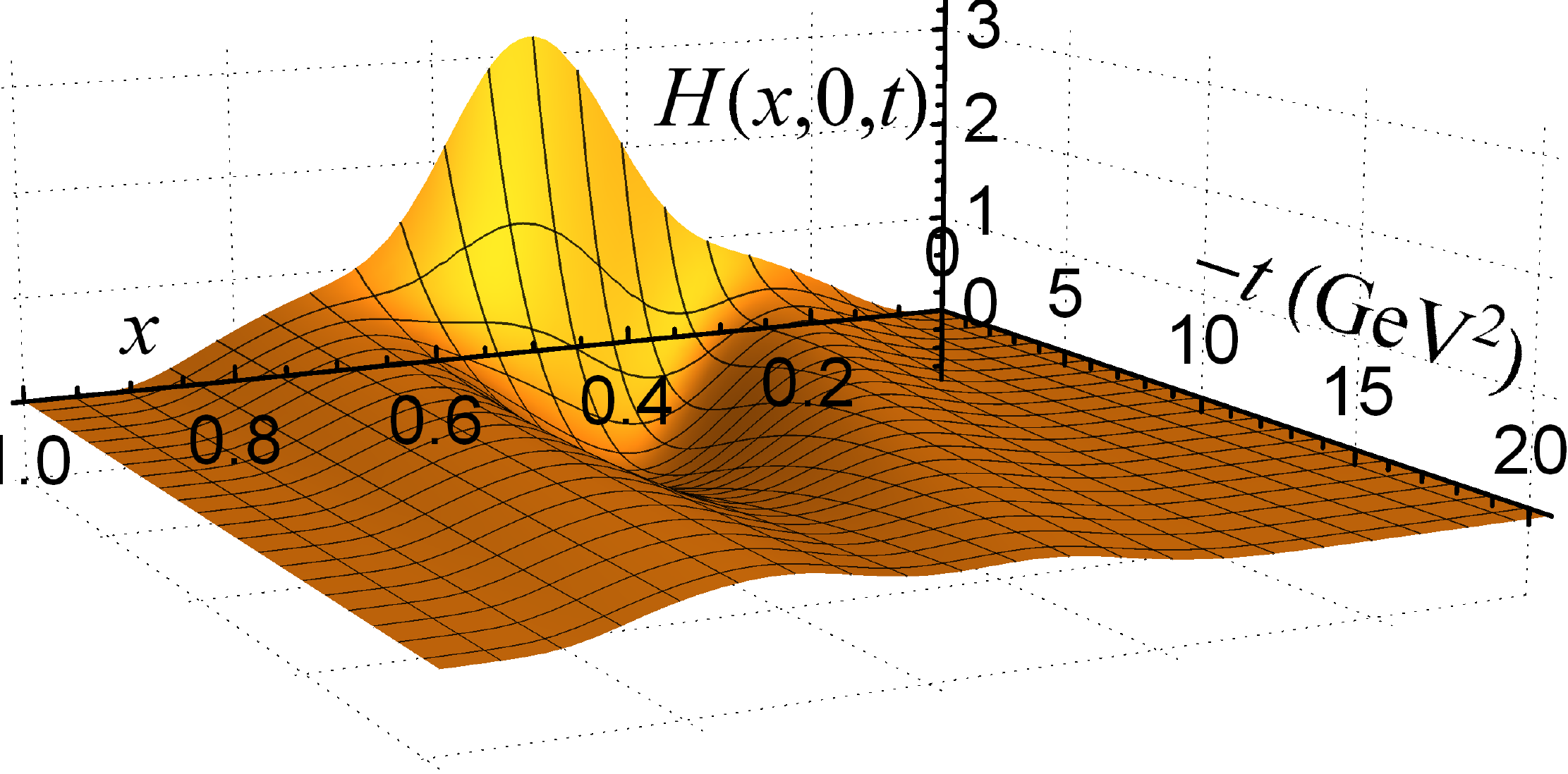}}\\ 
\subfloat[$\eta_{\mathrm{c}}^\prime: 2^1S_0 (0^{-+})$]{\includegraphics[scale=0.33]{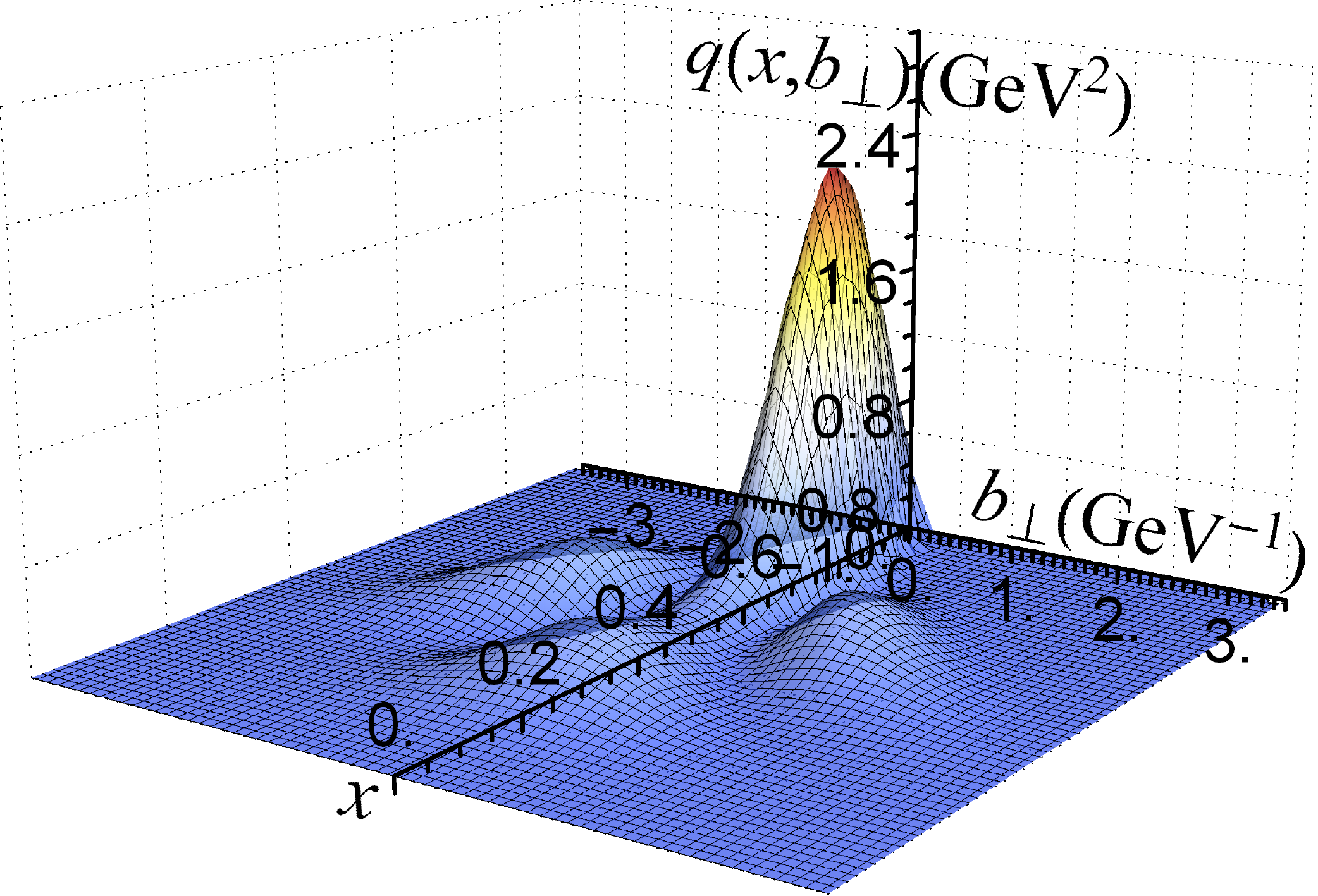}} \\ 
\end{tabular}
\caption{Top plots: 3D plot of helicity non-flip GPDs  $H(x, \xi=0, t=-\Delta_\perp^2)$ defined by Eq.~\ref{eq:gpds_intrinsic}. Bottom plots: impact-parameter dependent GPDs $q(x, b_\perp)$ defined by Eq.~\ref{eq:gpds_space}. 
Results are presented for $\mathrm{c}\overline{\mathrm{c}}$ bound states with $N_{\text{max}}=24$, $L_{\text{max}}=24$, $m_J=m_J' = 0$, coupling constant 
$\alpha=0.36$ (fixed coupling), the confining strength $\kappa = 0.94$ GeV, the 2D-HO basis scale $b =\kappa$, 
the quark (charmonium) mass $m_q =1.52$ GeV,  and gluon mass $\mu_g = 0.02$ GeV. 
Note that states are identified with their non-relativistic quantum numbers (relativistic quantum numbers) 
$n{}^{2S+1}\!L_J\, (J^{PC})$, 
where $n$ is the radial quantum number. The relation between $N$, the principal quantum number, 
and $n$, is $N=n+L$, where $L$ is the total orbital angular momentum, 
$S$ is the total intrinsic spin, $J$ is the total angular momentum,
$P$ is the parity and $C$ is the charge conjugation parity.}
\label{fig:gpd_b}
\end{figure}

To elucidate the structure of heavy quarkonium, we present 
an example of the range of observables accessible
with our LFWFs developed in Ref. \cite{Li:2015zda}.
From Ref.~\cite{Adhikari:2018}, we present in Fig. \ref{fig:gpd_b} the helicity-non-flip 
Generalized Parton Distribution (GPD)  $H(x, \xi=0, t=-{\vec \Delta}_\perp^2)$ 
for two bound states of charmonium. 
This GPD can be written, for the case where the photon couples only to the quark, 
as overlap integrals between LFWFs 
\cite{Brodsky:2000xy,Frederico:2009fk,Diehl:2003ny,Brodsky:2007hb,Adhikari:2016idg}:
\begin{equation}
H(x,\xi=0, t=-{\vec \Delta}_\perp^2) =  \sum_{\lambda_q,\lambda_{\bar q}}
\int d \vec k_\perp \, \psi^{J*}_{m_J^\prime}(\vec k'_\perp, x, \lambda_q,\lambda_{\bar q}) \psi^J_{m_J}(\vec k_\perp, x, \lambda_q,\lambda_{\bar q}). 
\label{eq:gpds_intrinsic}
\end{equation}
Here, $\bf k_\perp$ and $\bf k'_\perp $ are the respective relative transverse momenta of the quark 
before and after being struck by the virtual photon and $\Delta$ is the momentum transfer. We choose the Drell-Yan frame 
$\Delta^+ =0$, 
$t \equiv \Delta^2 = -{\vec \Delta}_\perp^2$. We also set skewness parameter to zero, $\xi=0$. 
And $\lambda_q (\lambda_{\bar {q}}) $ is the spin of the quark (antiquark).
The LFWF 
is normalized according to
$ \sum_{\lambda_q, \lambda_{\bar q}} \int dx \int d {\vec k}_{\perp} \big | \psi^J_{m_J}(\vec k_\perp, x, \lambda_q,\lambda_{\bar q}) \big |^2= 1$. 
These results represent demonstration cases illustrating hadronic GPDs in BLFQ 
similar to a positronium application in Refs.~\cite{Adhikari:2016idg,Vary:2016emi}.
 
Now, referring to Ref.~\cite{mb:GPD}, the impact-parameter dependent GPDs are defined as the Fourier transform of the GPDs with respect to the momentum transfer ${\vec \Delta}_\perp$:
\begin{equation}  
q(x, b_\perp) =
\int \frac{d{\vec \Delta}_\perp}{(2\pi)^2}
e^{-i {\vec \Delta}_\perp \cdot {\vec b}_\perp }
H(x,0,-\vec{\Delta}_\perp^2),
\label{eq:gpds_space} 
\end{equation}
which admits the partonic interpretation $\int dx \int d {\vec b}_{\perp} q(x, b_\perp)=1$.
Here, the impact parameter ${\vec b}_\perp$ corresponds to the displacement of the  quark $(q)$ from the transverse center of momentum of the entire $q\bar{q}$ system.  
We obtain the impact-parameter dependent GPDs $q(x, b_\perp)$ for the same two states of charmonium.  We present these GPDs in the lower panels 
of Fig. \ref{fig:gpd_b} to provide a visual impression of these coordinate space distributions. 
Note, especially, the appearance of secondary peaks.
We also note that these impact-parameter dependent GPDs $q(x,b_\perp)$ are not symmetric 
with respect to $x=1/2$, as $\vec b_\perp =(1-x) {\vec r_\perp}$ is conjugate to $\vec \Delta_\perp$. 

\begin{figure}
 \centering 
\includegraphics[width=1.00\textwidth]{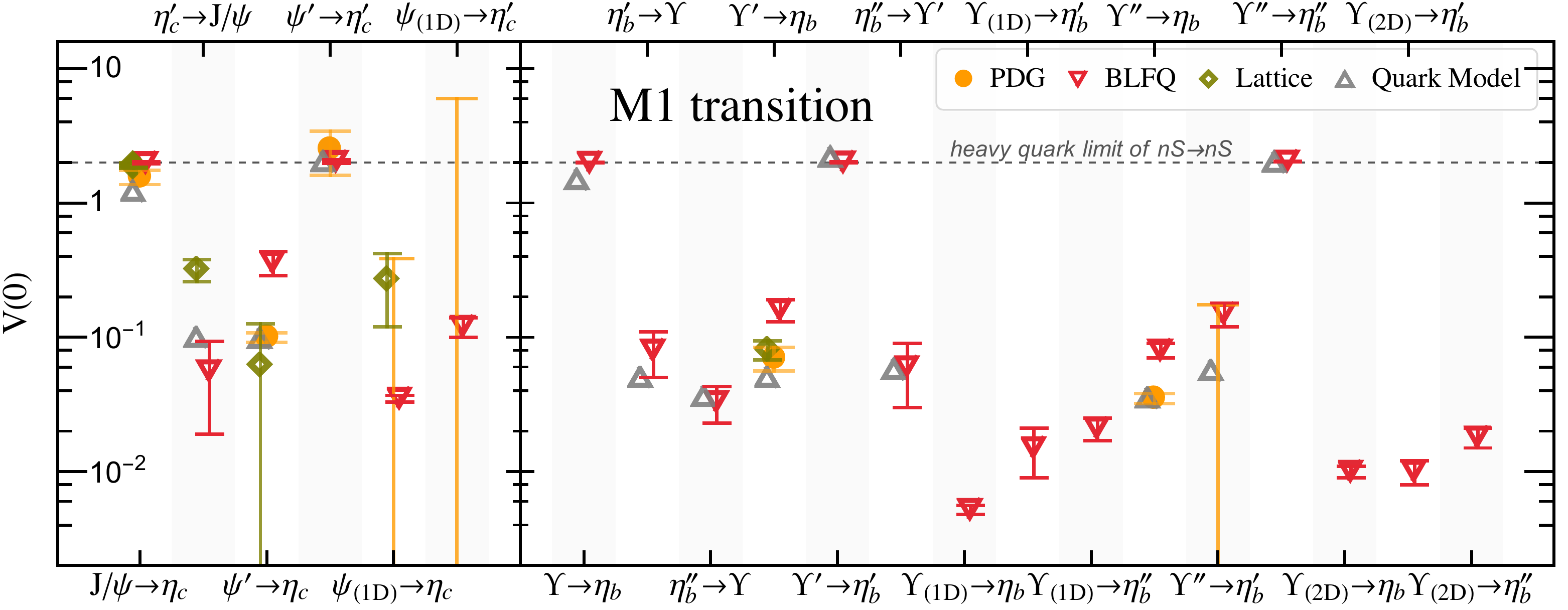}
\caption{(Color online) Transition form factors $V(0)$ of charmonia and bottomonia transitions. 
Points are based on extrapolations using $N_{\rm max}$ = $L_{\rm max}$ = 8, 16, 24 and 32 for fits with 
second-order polynomials in $N_{\rm max}^{-1}$. We use the difference between the extrapolated and the
$N_{\rm max}$ = 32 result to quantify the uncertainty which is shown as an error bar.  
The labels on the horizontal axis define the transitions with a prime signifying a 2$S$ state and a double prime
signifying a 3$S$ state. Experimental results (PDG) are from Ref. \cite{Patrignani:2016xqp} 
while Lattice QCD results are from Refs. \cite{Dudek:2006ej,Dudek:2009kk,Becirevic:2014rda,Hughes:2015dba}. The quark model results are from Ref.~\cite{Ebert:2002pp}.}
\label{fig:decay_V}
\end{figure}

Building on these achievements, we employ the LFWFs of Ref.~\cite{Li:2017mlw} to evaluate the radiative
transitions between vector and pseudo-scalar states of heavy quarkonium  \cite{Li:2018aaa}. These magnetic dipole (M1) transitions require the emission of a photon from either the quark or antiquark with a spin-flip. For the vector to pseudo-scalar transition, the decay width is expressed as:
\begin{equation}
 \Gamma(\mathcal{V} \to   \mathcal{P}\gamma)
=
    \frac{ \alpha_{EM} {(m_{\mathcal{V} }^2-m_{\mathcal{P}}^2)}^3}{ {(2m_{\mathcal{V}})}^3{(m_{\mathcal{P}}+m_{\mathcal{V}
        })}^2}\frac{4}{3}\mathcal{Q}_f^2{|V(0)|}^2
\label{eq:rad_width} 
\end{equation}
where $V(0)$ is the transition form factor extracted from the matrix element of the hadronic current operator $J^\mu$ at zero momentum transfer. We evaluate the transition form factors with the transverse current and the longitudinally polarized state ($m_j=0$) of vector mesons. 
We present our results for $V(0)$ in Fig.~\ref{fig:decay_V} using the LFWFs of Ref.~\cite{Li:2017mlw} for states below their respective open flavor thresholds. Considering the residual sensitivity to the choice of current operator due to Fock space truncation, 
we find reasonable agreement with experiment~\cite{Patrignani:2016xqp}, with lattice QCD  
\cite{Dudek:2006ej,Dudek:2009kk,Becirevic:2014rda,Hughes:2015dba} and with quark model~\cite{Ebert:2002pp}.

In closing this section, we compare in Table I the root mean square (rms) radii of the  $\eta_{c}$ \& $J/\psi$ mesons
between the BLFQ approaches discussed above and an earlier treatment of the Bethe-Salpeter approach
referred to as the Variational Tamm-Dancoff (VTD) approach~\cite{Spence_etal}. In the current application of the VTD approach,
the equations are rewritten in light-front coordinates to facilitate more extended comparisons with the BLFQ 
approach.  Note that the VTD results in Table I are within a few percent of the BLFQ results where both approaches employ the running coupling constant.
\begin{table}
\begin{centering}
\begin{tabular}{|c|c|c|c|c|}
\cline{3-5} 
\multicolumn{1}{c}{} &  & VTD - running coupling~\cite{Spence_etal} & BLFQ - running coupling~\cite{Li:2017mlw} & BLFQ - fixed coupling~\cite{Li:2015zda}\tabularnewline
\hline 
\hline 
\multirow{2}{*}{rms radius (fm)} & $\eta_{c}$ & 0.179 & 0.170 & 0.199\tabularnewline
\cline{2-5} 
 & $J/\psi$ & 0.172 & 0.175 & 0.212 \tabularnewline
\hline 
\end{tabular}
\par\end{centering}
\caption{Ground state root mean square (rms) radii for the $\eta_{c}$ \& $J/\psi$ mesons
within different approaches. The values shown for Variational Tamm Dancoff (VTD) \cite{Spence_etal}
and BLFQ with running coupling were evaluated with an $N_{\max}=32$ basis~\cite{Li:2017mlw},
while the BLFQ with fixed coupling results~\cite{Li:2015zda} were evaluated with an
$N_{\max}=24$ basis. }
\end{table}
\section{Mixed Flavor Heavy Quarkonium} \label{sec 5}
Following up on the successful applications of BLFQ to single-flavor heavy quarkonium \cite{Li:2015zda,Li:2017mlw}, 
we investigate an example of the unequal mass relativistic bound state system, the $B_c$ mesons, with the Hamiltonian of
Eq.~\ref{eq:effective_hamiltonian} again solved in the BLFQ approach \cite{Tang:2018bbb}. 
Following Ref. \cite{Li:2017mlw} we employ its running coupling for the one-gluon exchange, 
retain its fitted values of the quark masses, and
adopt the mean-square average of the charmonium and bottomonium results 
for the confining strength $\kappa$. The direct adoption of these parameters produces the 
ground state of $B_c$ at 6.258 GeV which is close to the experimental result of 6.2749(8) GeV 
\cite{Patrignani:2016xqp} and the lattice result of 6.304(12) GeV~\cite{Allison:2004be}.
We use the LFWFs for the $B_c$ mesons to calculate decay constants.  We find reasonable agreement among our results \cite{Tang:2018bbb}, experimental results \cite{Patrignani:2016xqp} and other theoretical approaches \cite{AbdElHady:1998kc,Choi:2009ai,Colquhoun:2015oha,Koponen:2017fvm}.
\section{Light Mesons} \label{sec 6}
\begin{figure}[ht]
\centering
\begin{minipage}[b]{0.475\linewidth}
\includegraphics[width=0.98\textwidth,clip]{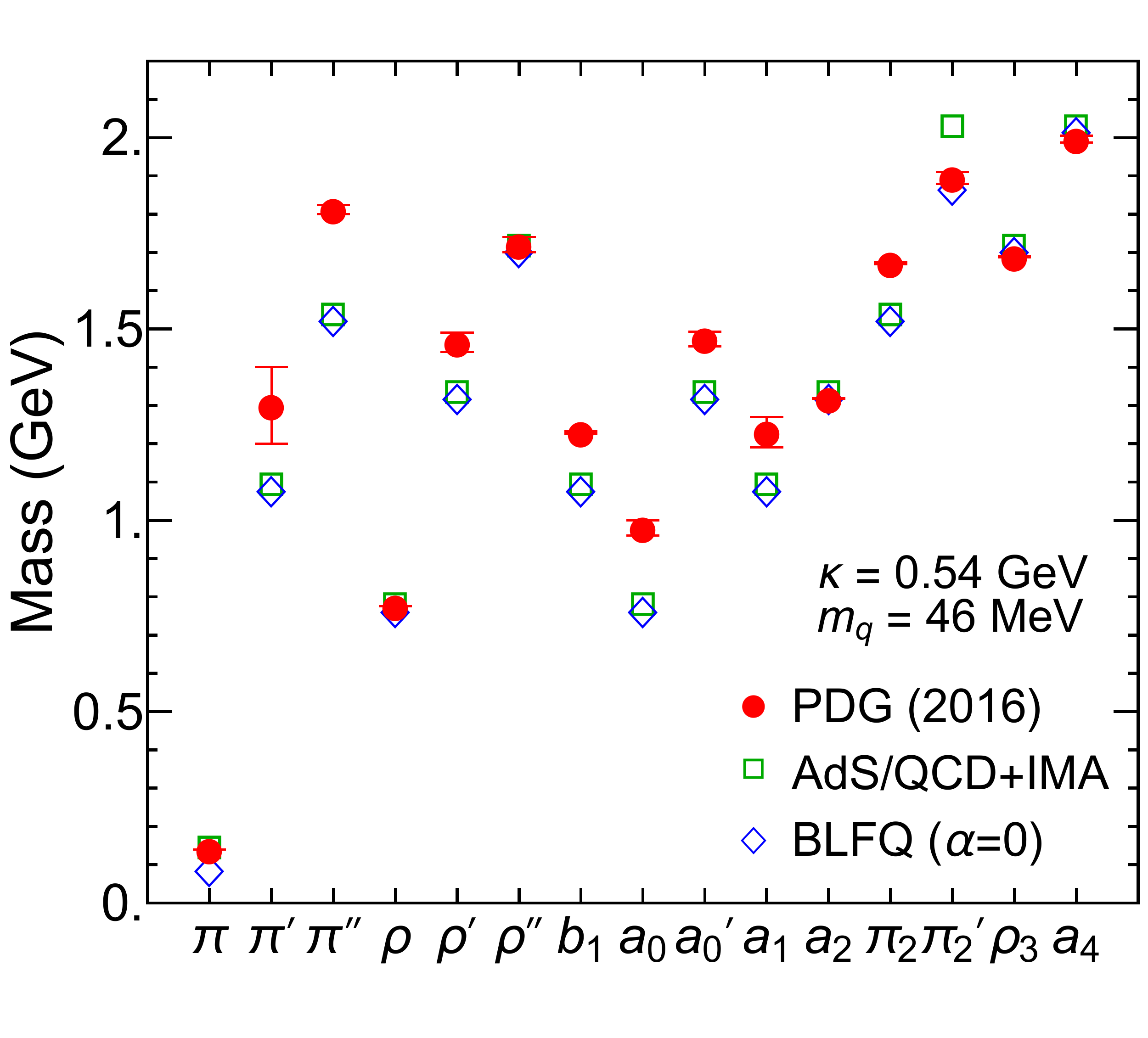}\\[2mm]
\end{minipage}
\begin{minipage}[b]{0.515\linewidth}
\includegraphics[width=0.98\textwidth,clip]{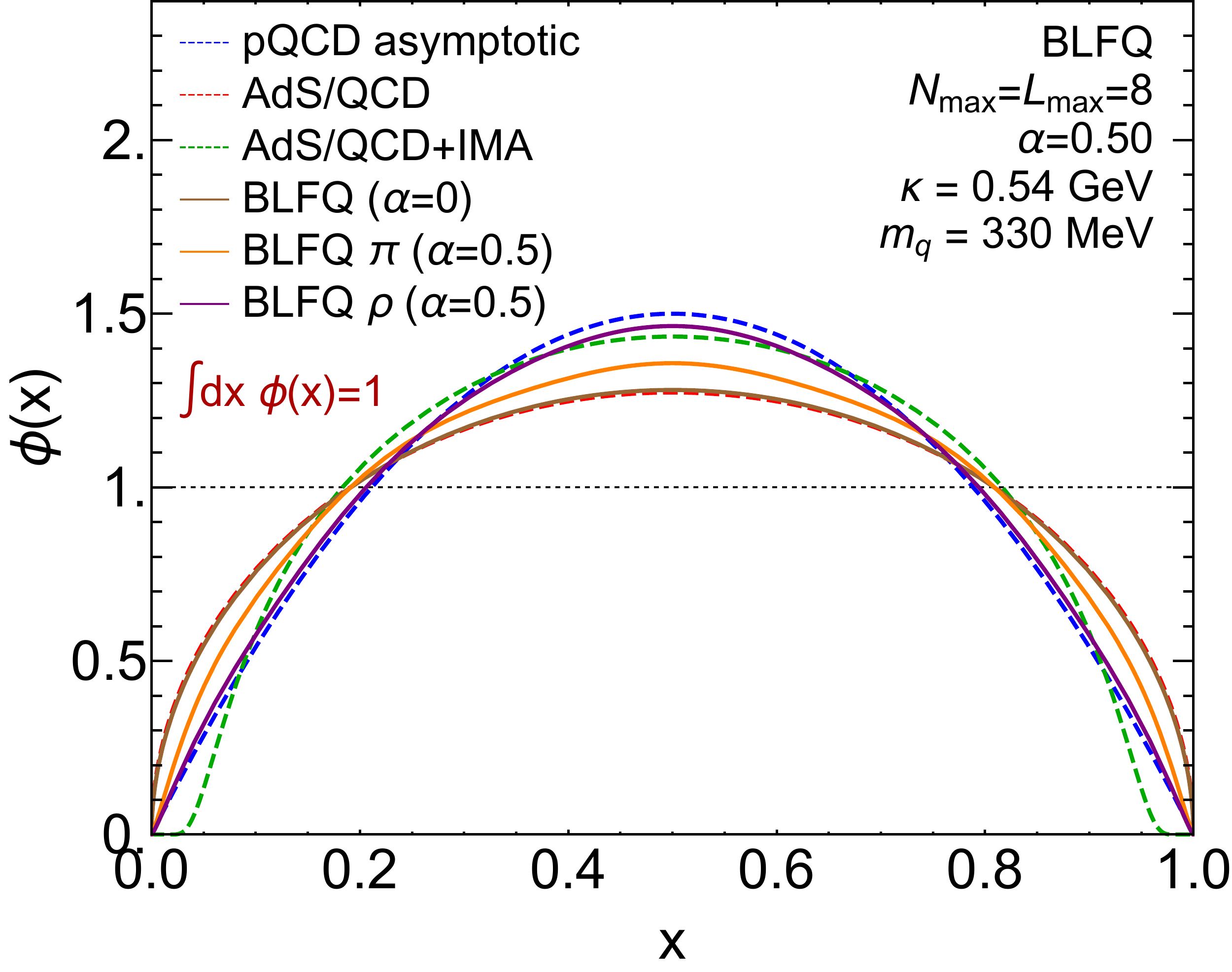}\\[2mm]
\end{minipage}
\caption{\label{fig:light_mesons}  (Color online) Comparisons of the mass spectra (left panel)
and the quark distribution amplitudes for the $\rho$ meson (right panel) obtained from various
models. Experimental masses come from PDG~\cite{Patrignani:2016xqp}.
See the references in the text for the definitions of various models.
}
\end{figure}
Before including the contributions from the one-gluon exchange, it is instructive
to compare results of existing light-front models for light mesons to available experimental data. For this purpose, we select the model of AdS/QCD \cite{Brodsky:2014yha} 
and AdS/QCD augmented by the invariant mass ansatz in Ref.~\cite{Brodsky:2014yha} as well as our own model of 
Eq.~\ref{eq:BLFQ_effective_hamiltonian} with the one-gluon exchange 
term suppressed ($\alpha$=0).
For these models we present comparisons of the mass spectra 
and $\rho$ meson distribution amplitudes (DAs) in Fig. \ref{fig:light_mesons} excerpted 
from Ref. \cite{Qian:2018aaa}. 
Both models on the left panel of Fig.~\ref{fig:light_mesons} provide remarkably similar mass spectra.
They also represent the general features of the experimental spectra with similar tendencies to produce lighter excited states.

\begin{figure}
\centering
\includegraphics[width=0.6\linewidth]{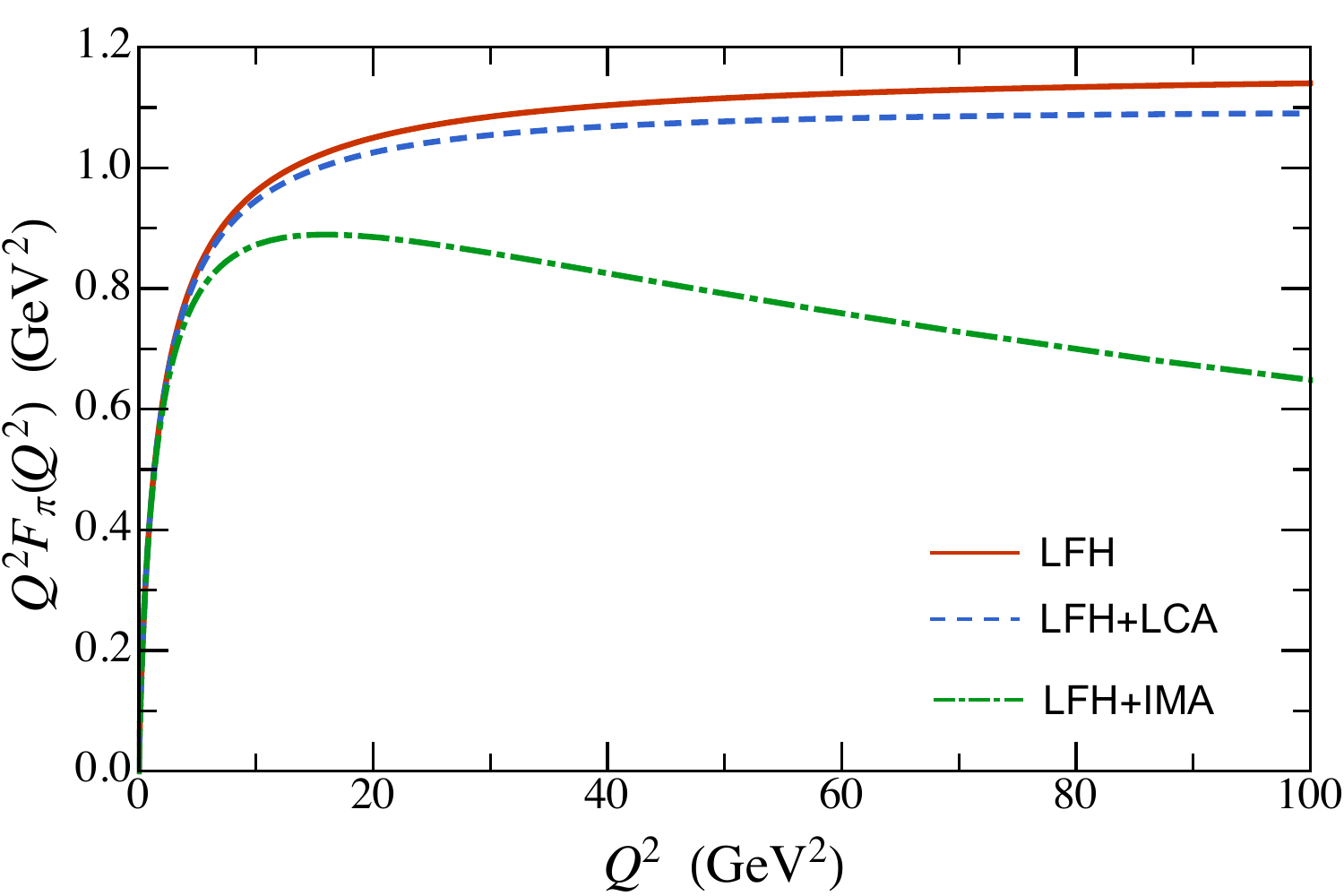}
\caption{Comparison of the pion form factor calculated from the light front holography (LFH), LFH with LCA, and LFH with IMA.}
\label{fig:pion_UV_asymptotics}
\end{figure}

The quark DAs are defined from the light cone correlators for the valence quark LFWF. For models discussed here in the 
$\ket{q\bar{q}}$ Fock sector, the 
quark DAs are equivalent to the valence sector LFWF in coordinate space at $\mathbf{r}_\perp = 0$. The right panel of Fig.~\ref{fig:light_mesons} shows the comparison of the $\rho$ meson DAs from the BLFQ, ADS/QCD and perturbative QCD (pQCD).
We note the significant differences in the width of these distributions and their behaviors near the end points. The large difference between the AdS/QCD and AdS/QCD+IMA models reveals the major effect of nonvanishing quark masses at the DA end points. Specifically, the pion form factor from AdS/QCD+IMA has the unphysical UV asymptotics. Whereas using AdS/QCD with the longitudinal confinement ansatz (LCA), the correct UV asymptotics is retained. See Fig.~\ref{fig:pion_UV_asymptotics} for details.

We find that the inclusion of the one-gluon exchange contribution in 
Eq. \ref{eq:BLFQ_effective_hamiltonian} (i.e. BLFQ with all terms included) 
has a large effect for the lightest mesons.  For example, 
without a source for chiral symmetry breaking, one anticipates it is difficult
to obtain the experimental $\pi - \rho$ mass splitting. To remedy this deficiency,
we plan to introduce a phenomenological pseudo-scalar $q \bar q$ interaction.
Preliminary results indicate that a pseudo-scalar 
interaction added to the light-front Hamiltonian of 
Eq. \ref{eq:BLFQ_effective_hamiltonian} 
provides a good fit to the mass spectra. Tests of
additional meson properties with this augmented Hamiltonian are underway.
\section{Baryon Systems} \label{sec 7}
\begin{figure}[ht]
\centering
\begin{minipage}[b]{0.41\linewidth}
\includegraphics[width=0.98\textwidth,clip]{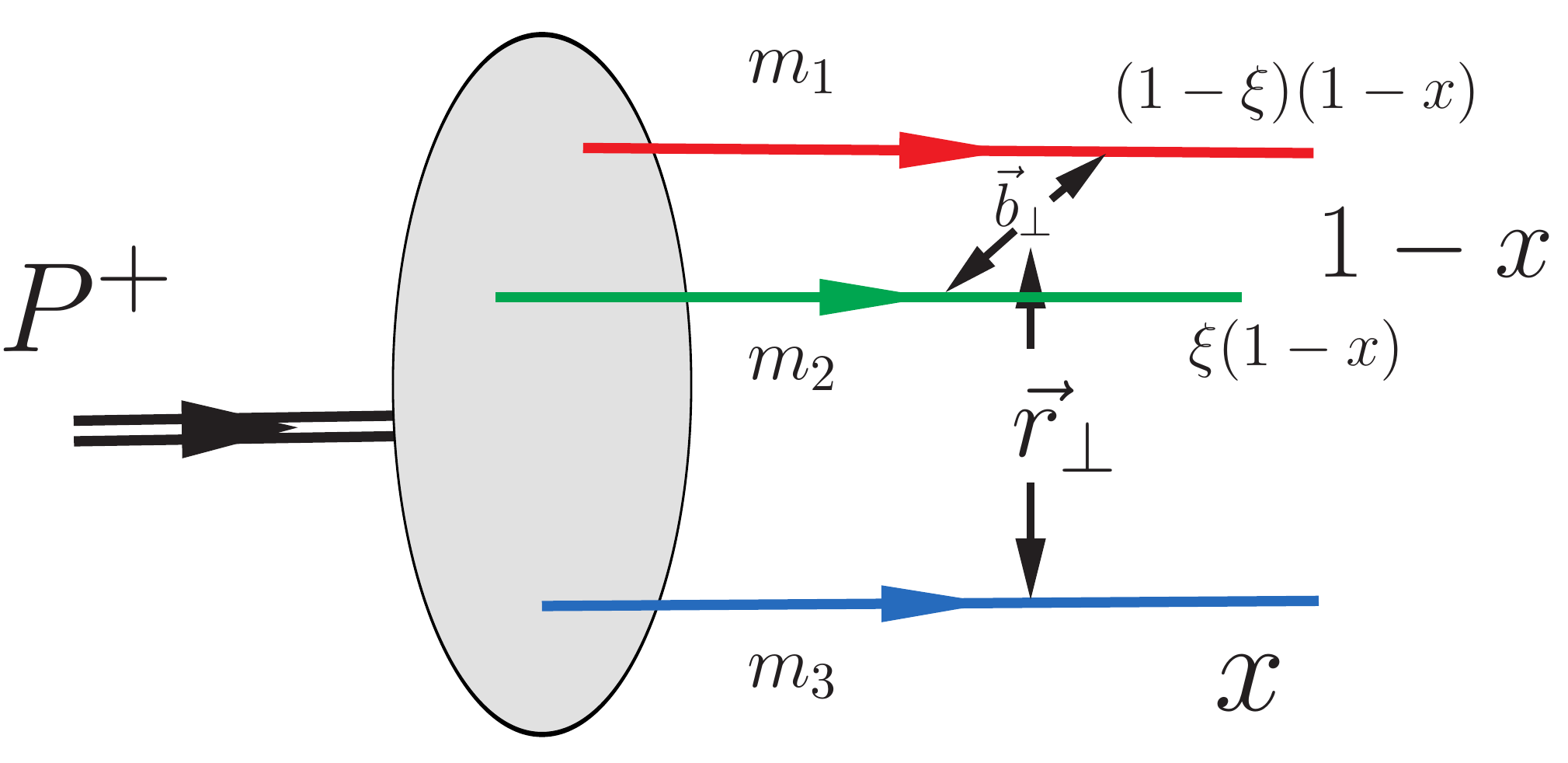}\\[2mm]
\includegraphics[width=0.98\textwidth,clip]{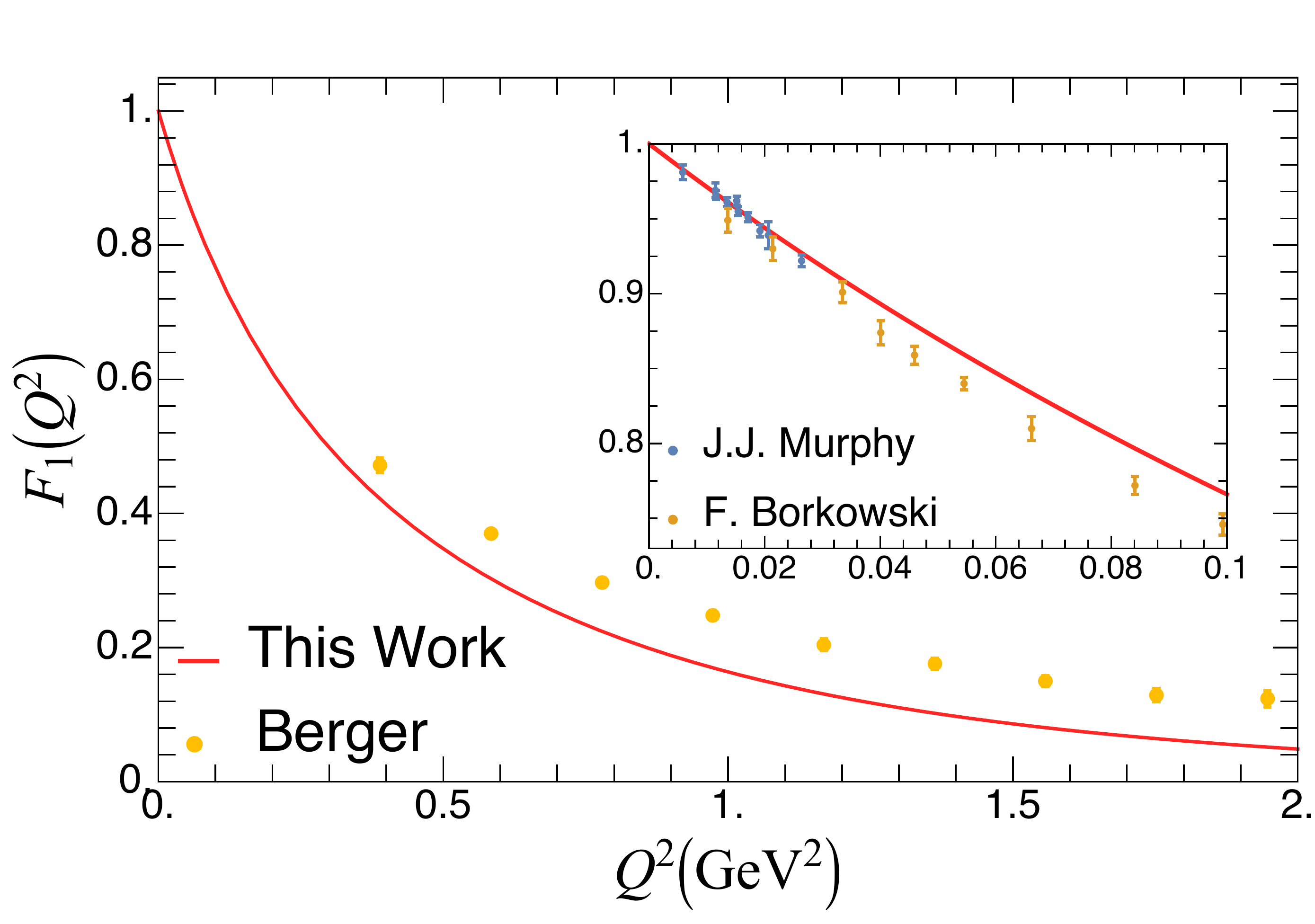}\\[2mm]
\end{minipage}
\begin{minipage}[b]{0.505\linewidth}
\includegraphics[width=0.98\textwidth,clip]{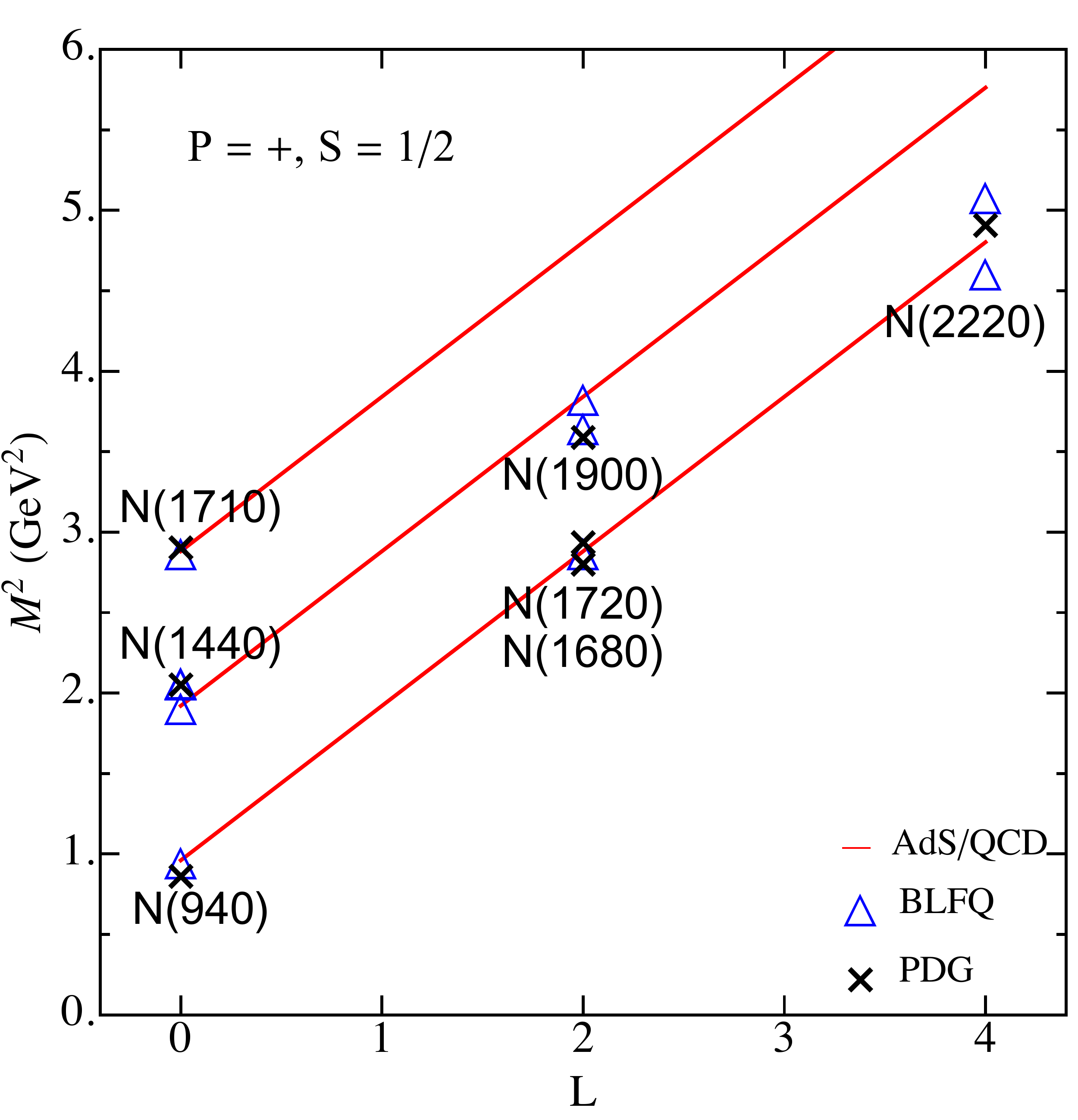}\\[2mm]
\end{minipage}
\caption{\label{fig:BLFQ_baryon}  (Color online) 
Top left panel: Sketch showing the Jacobi coordinates for a baryon with three quarks
with masses ${\rm m_j}$. 
Right panel: Comparisons of the mass-squared spectra 
in GeV$^2$ as a function of the angular momentum ${\rm L}$ assigned in AdS/QCD~\cite{Brodsky:2014yha}. 
Experimental masses (PDG) are from Ref. \cite{Patrignani:2016xqp}.
Lower left panel: The proton charge form factor from BLFQ as a function 
of $Q^2$ in GeV$^2$ and compared with experiments~\cite{Borkowski:1974mb,Murphy:1974zz,Berger:1971kr}.
}
\end{figure}

Following the promising applications of BLFQ to QED and to the
meson sector of QCD, we embark upon an application to the 
baryon systems. We carry out a direct generalization of 
Eq. \ref{eq:BLFQ_effective_hamiltonian} to the three
unequal quark mass system using Jacobi coordinates
explained in the top left panel of Fig. \ref{fig:BLFQ_baryon}
\cite{Yu:2018aaa}.
This generalization yields mass eigenvalues of the form 
\begin{eqnarray}
M^2_{\mathbf{n}_1,\mathbf{m}_1,\mathbf{n}_2,\mathbf{m}_2,L,l} = ({\rm m_3}+{\rm M}_L)^2+2\kappa^2(2\mathbf{n}_1+|\mathbf{m}_1|+2\mathbf{n}_2+|\mathbf{m}_2|+2) \nonumber \\
+\frac{{\rm M}_L+{\rm m_3}}{{\rm m_1+m_2+m_3}}\kappa^2(2l+1)
+\frac{\kappa^4}{({\rm m_1+m_2+m_3})^2 }l(l+1) + {\rm const.},\\
{\rm M}_L^2 = ({\rm m_1+m_2})^2 + \frac{{\rm m_1+m_2}}{{\rm m_1+m_2+m_3}}\kappa^2(2L+1)
+\frac{\kappa^4}{({\rm m_1+m_2+m_3})^2 }L(L+1)
\label{eq:BLFQ_Baryon_mass}
\label{baryon_mass2}
\end{eqnarray}
where $\mathbf{n}_i$ ($\mathbf{m}_i$) are the 2D-HO radial (orbital angular momentum projection) quantum numbers, $l$ ($L$) is the longitudinal quantum number and ${\rm m_j}$ are the quark masses. 

In the right panel of Fig. \ref{fig:BLFQ_baryon}, we present results of the solved 
BLFQ spectra of Eq. \ref{baryon_mass2} for the positive parity, spin $1/2$ baryons to compare with the results of AdS/QCD \cite{Brodsky:2014yha} and with
experiment.  The results are presented as a function of ${\rm L}$, the angular momentum
quantum number defined in AdS/QCD \cite{Brodsky:2014yha}. For the BLFQ results, 
we employ equal quark masses ${\rm m_j}$ = 0.35 GeV, $\kappa=0.49$ GeV and zero for the 
longitudinal quantum numbers $L$ and $l$. 
Overall, there is good agreement between AdS/QCD and BLFQ as well as good agreement between these models
and experiment. In particular this signals the appearance of Regge trajectories for the baryons in
this simplified BLFQ approach, not significantly modified by the longitudinal confinement potential and the appearance of three dynamical quarks.

Additional observables may be accessed within this BLFQ model of the baryons.  As an example, 
we display the proton charge form factor from BLFQ as a function of $Q^2$ in GeV$^2$ and compared with experiments \cite{Borkowski:1974mb,Murphy:1974zz,Berger:1971kr} in the lower
left panel of Fig. \ref{fig:BLFQ_baryon}.  There is good agreement at lower $Q^2$ which is 
controlled by choosing $\kappa=0.49$ GeV, a value that produces $0.9$ fm for the proton charge radius
in the BLFQ model, which is close to experiment. At sufficiently high $Q^2$, we expect the BLFQ result to fall below the experimental charge form factor as observed in Fig.~\ref{fig:BLFQ_baryon}.
We anticipate that including the one-gluon exchange contribution in BLFQ and diagonalizing
the resulting effective Hamiltonian in a sufficiently large basis will improve the proton charge
form factor at higher $Q^2$~\cite{Yu:2018aaa}.
\section{Quark Jet Scattering in a Color Glass Condensate} \label{sec 8}
We are working towards a time-dependent treatment of high-energy reactions dominated by both the intense electromagnetic fields and the strong interaction. Specifically, we have adopted the tBLFQ framework \cite{Zhao:2013jia,Zhao:2013cma} to investigate the effects of electromagnetic (EM) fields generated by ultra-relativistic heavy ions on charged particles \cite{Chen:2017uuq,Vary:2016ccz}. This study is motivated by the fact that strong EM fields are generated during heavy ion collisions \cite{Tuchin:2013ie}, and a quantitative study of their effects on charged particles is essential for extracting  properties of the Quark Gluon Plasma (QGP) as well as the properties of the gluon distributions in nuclei. 
\begin{figure}
 \centering 
\includegraphics[width=0.35\textwidth,scale=1.2]{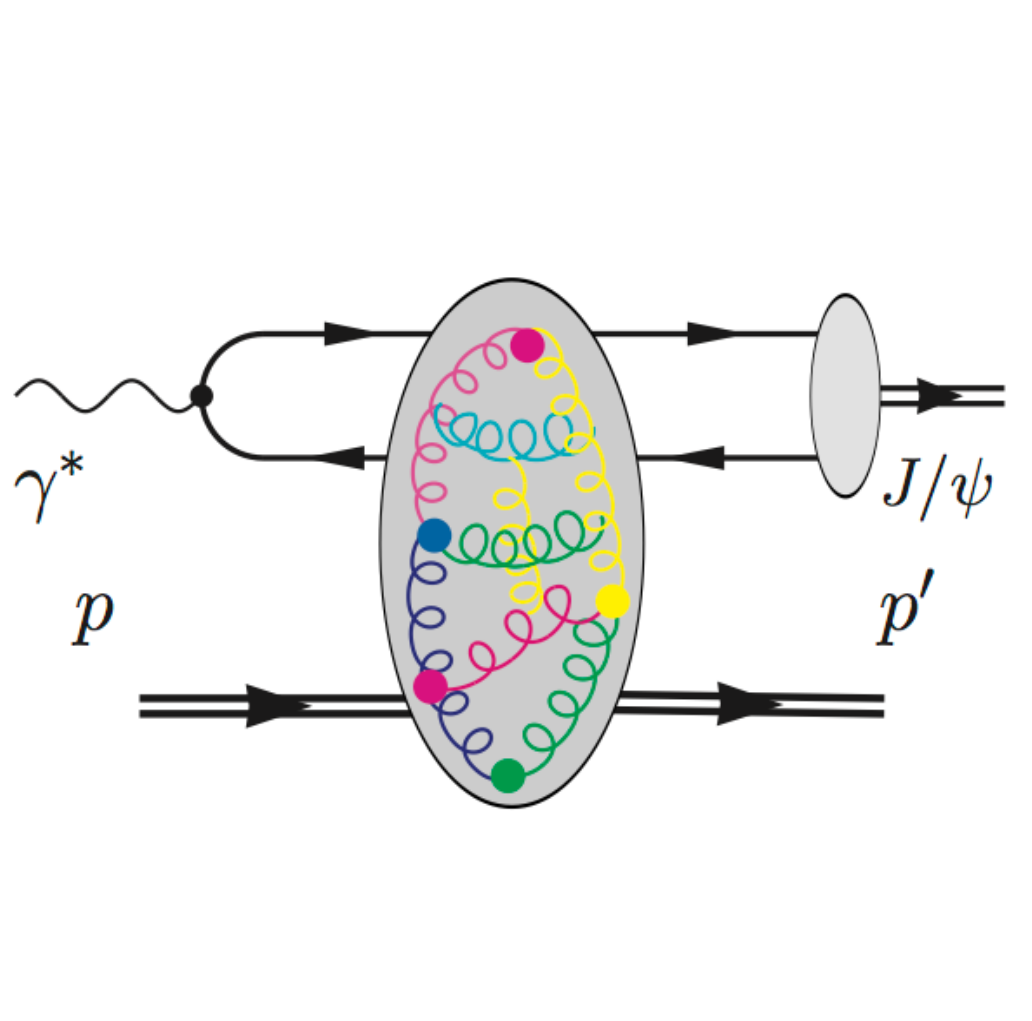}
\includegraphics[width=0.35\textwidth]{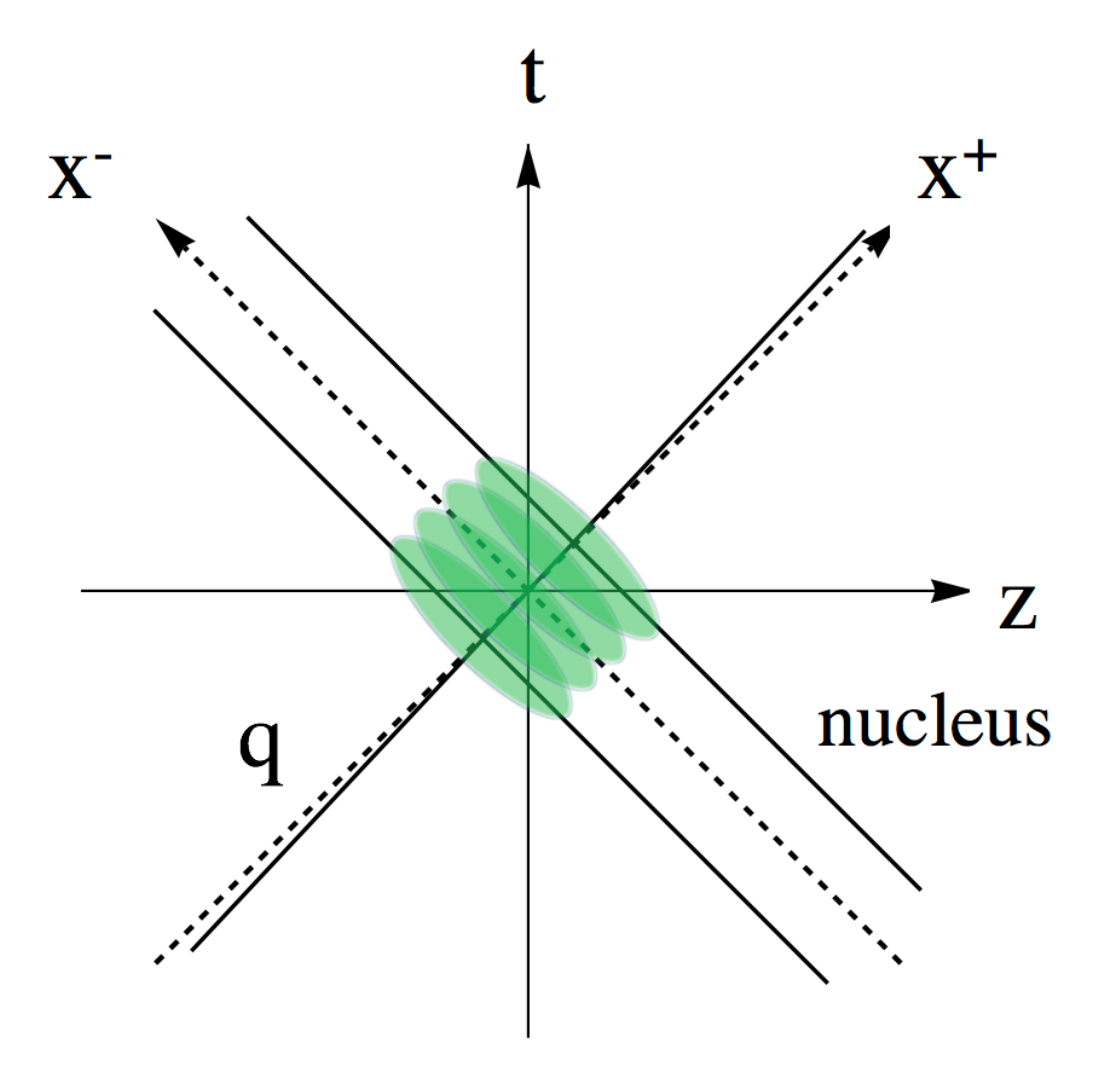}
\includegraphics[width=0.55\textwidth]{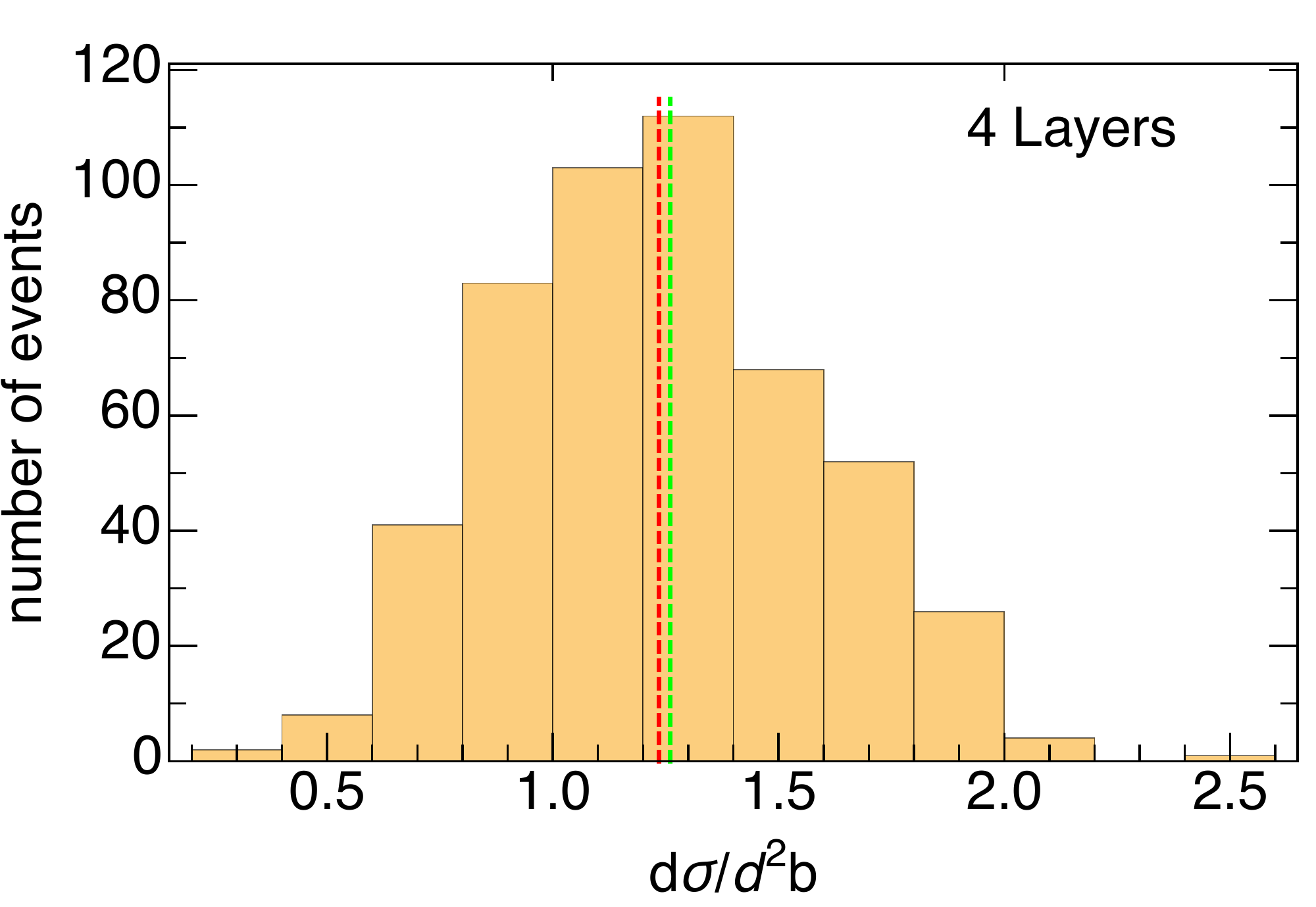}
\caption{(Color online)  
\textit{Top Left}: Schematic of vector meson production in the dipole model 
that begins with virtual photon dissociation to a $q \bar q$ pair, interaction with the gluon field 
of a hadron with 4-momentum $p$ and hadronization of the vector meson in the final state.
\textit{Top Right}: Sketch of quark scattering on color glass condensate in 2D light-front coordinates. 
\textit{Bottom}: Distribution of events for quark scattering four layers of a U(1) color glass condensate
with $g^2 \mu$, the color charge density,  is 0.5 GeV and $\Lambda$, the infrared scale,  is 0.1 GeV.
The vertical green line is an analytical result following Ref.  \cite{Dumitru:2002qt} and the vertical
red line is the average over 500 events, each obtained from an independent initial gluon field.}
\label{CGC_scattering}
\end{figure}

We have investigated the real-time evolution of a quantized fermion field (taken initially as an electron) 
under the influence of a strong external time-dependent EM field of a relativistic heavy ion \cite{Chen:2017uuq}.
When the coupling between the electron and the external field is as small as the field generated by an ultra-relativistic proton with the physical coupling $ 
\alpha_\text{em} \approx 1/137$, the transition rate between two kinetic energy eigenstates of the electron calculated within the tBLFQ approach shows agreement with light front perturbation theory (LFPT). In contrast, for EM fields generated by an ultra-relativistic gold nucleus, the coupling of an electron to the fields is $Z_\text{Au}\alpha_\text{em} \approx 79/137$. The transition rate between the same two kinetic energy eigenstates calculated within the tBLFQ approach deviates from LFPT calculations (both leading-order and next-to-leading order), and the differences among these results provide an indication of the significance of higher order perturbation effects~\cite{Chen:2017uuq}. 

Since there are good prospects for applying the tBLFQ formalism to heavy ion collisions and electron ion collisions, we are motivated to aim for more realistic applications. For instance, we have adopted the classical description of gluon fields in high energy nuclear collisions from the Color Glass Condensate (CGC) effective theory \cite{McLerran:1993ni} to eventually investigate the real-time evolution of colored objects interacting with these classical gluon 
fields~\cite{Li:2018bbb}. 
Within the tBLFQ framework, we can study the effects of gluon fields generated in the initial stage of relativistic heavy-ion collisions on heavy quarks and jets. The advantages of the tBLFQ framework are distinctive: the calculation is both relativistic and at the amplitude level thereby incorporating quantum interference effects. In addition, we can naturally extend our calculation to higher Fock sectors and go beyond the Eikonal approximation.        

For an initial application, we consider a quark propagating in the color glass condensate~\cite{Li:2018bbb} as depicted  
occurring during the intermediate stage of vector meson production in the left panel of Fig. \ref{CGC_scattering}. 
In tBLFQ, the quark configurations are described by the BLFQ basis, which is a representation of localized but freely propagating quarks.  We evolve the initial state (one chosen eigenstate) 
over light-front time under the influence of the color glass condensate (a U(1) field in this demonstration).  The gluon field is obtained by numerically solving the Yang-Mills equations with a stochastic color source of charge density $g^2\mu$. 
The scattering is treated here as occurring through 4 layers 
of the color glass condensate and 500 cases are time-evolved to generate the distribution of cross sections 
shown in the bottom panel of Fig. \ref{CGC_scattering}. The average over these events is in good agreement with
the analytical solution \cite{Dumitru:2002qt} as one expects.  This signals pathways to include additional layers
and to move forward to the case of the SU(3) chromodynamic field. 
\section{Glueball Sector} \label{sec 9}
\begin{figure}[ht]
\centering
\begin{minipage}[b]{0.475\linewidth}
\includegraphics[width=0.98\textwidth,clip]{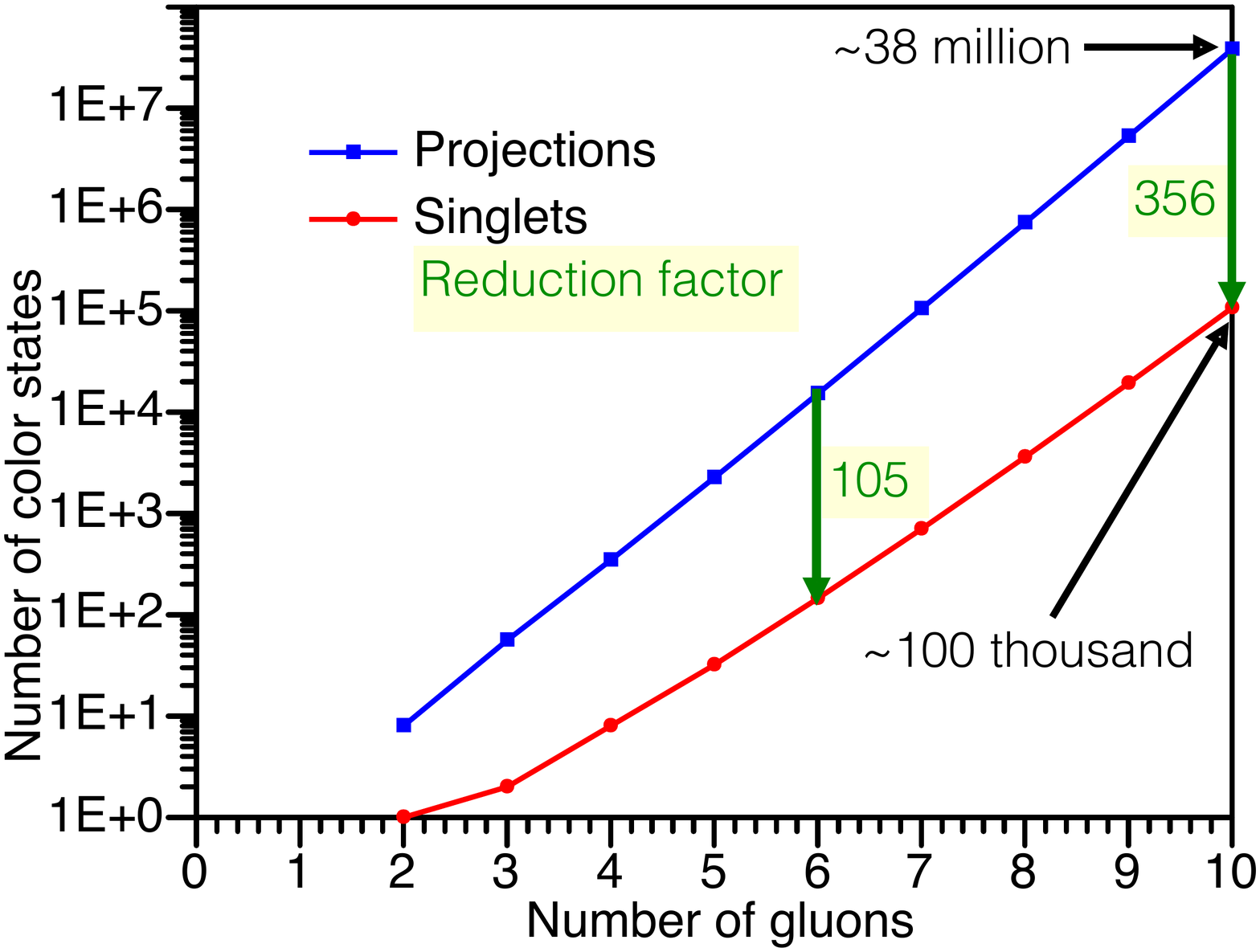}\\[2mm]
\end{minipage}
\begin{minipage}[b]{0.515\linewidth}
\includegraphics[width=0.98\textwidth,clip]{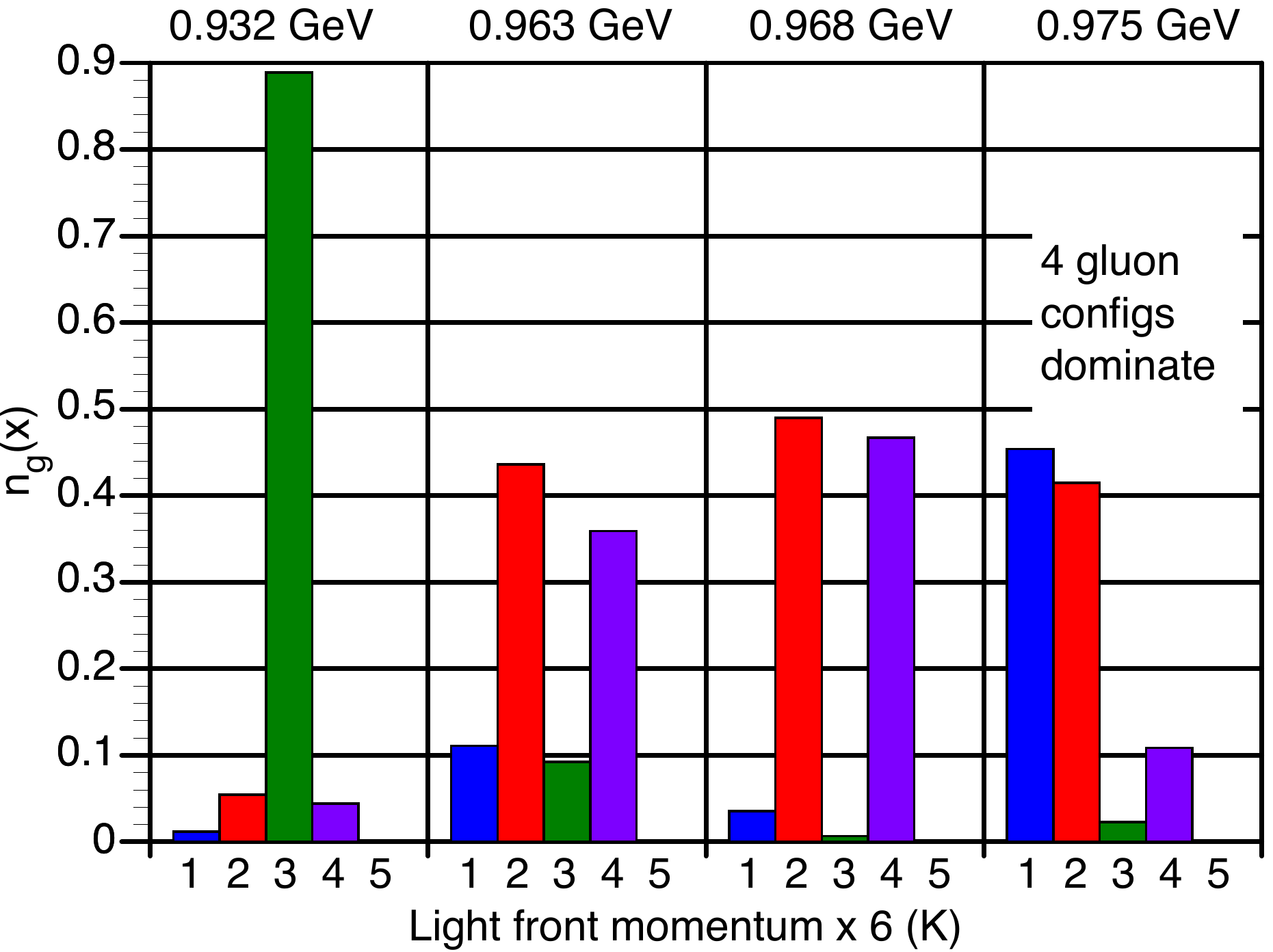}\\[2mm]
\end{minipage}
\caption{\label{fig:glueball}  (Color online) {\it Left:} Comparisons of the matrix dimensions for basis spaces in each Fock sector with color singlet projection = 0 (``Projections'') and with basis spaces having
only total color singlets (``Singlets'') as functions of the number of gluons.
This is an extended version of the results in Ref.~\cite{Vary:2009gt}.  {\it Right:} Gluon 
light-front momentum distribution fractions $n_g(x)$ for each of the lowest 4 mass 
eigenstates  in the $N_{\rm max} = L_{\rm max} = 6$ calculation discussed in the text.  
The heading of each column presents the mass of the glueball. 
}
\end{figure}
Moving from pure fermionic systems with effective interactions for
the gluon degrees of freedom, we turn to the pure glue sector as a
first application of BLFQ where we eliminate the restrictions of an isolated Fock sector truncation.  Here,
we will neglect the gluon zero modes and adopt
discretized plane waves with periodic boundary conditions (DLCQ) 
for the longitudinal modes.
We then retain all the gluon Fock sectors consistent with the 
BLFQ regulators $N_{\rm max}$ and total longitudinal momentum $K$.
This treatment of the longitudinal modes, including all Fock sectors consistent
with the regulator $K$, reflects earlier applications
to 1+1 dimensional scalar field theories \cite{Harindranath:1987db}.
Higher Fock sectors in scalar boson theories were found to be important
to map out the topological phases, kinks and anti-kinks,  
as coherent solutions in the Hamiltonian light-front
quantization approach~\cite{Harindranath:1987ex,Chakrabarti:2003ha,Chakrabarti:2003tc,Chakrabarti:2005zy}.

Due to the rapid increase in basis space with increasing regulators, we will restrict the initial application to $N_{\rm max} = L_{\rm max} = 6$. Our Fock space then includes states with 2 through 6 gluons. Taking 
into consideration the restriction to gluon color singlet configurations,
we obtain a Hamiltonian matrix dimension of 1987.  The importance
of restricting the basis to global color singlets rather than using
color configurations with zero color projection can be easily seen in the left panel of Fig. \ref{fig:glueball}.  Here, the dimension of the gluon
color configurations in each Fock sector is shown to be
reduced by up to two orders of magnitude by adopting
color singlet configurations rather than the zero color projection 
configurations.  The vertical arrow in Fig. \ref{fig:glueball} highlights
that reduction for the Fock space states with 6 gluons.
With this color singlet basis, the BLFQ glueball spectra and LFWFs 
are calculable in about 3 minutes on a laptop.

For the effective multi-gluon Hamiltonian we will employ simply the 
kinetic energies of the ``constituent'' gluons, each with mass 
$m_g= 0.25$ GeV, and the triple-gluon coupling with strength $g_s = 0.5$.  For the 2D-HO basis parameter we choose $b = 0.50$ GeV. 
The instantaneous gluon exchange term and the 
quartic gluon coupling (both second order in $g_s$) will be 
included at a later stage.  

One should consider the specific mass eigenvalues as model results
that are dependent on the values of the model parameters. It is intriguing to observe the different features of the gluon distribution functions
$n_g(x)$ for each of the lowest 4 masses as shown in the
right panel of Fig.~\ref{fig:glueball}. The lowest 3 states are dominated
by 2-gluon configurations but with distinctive distributions.  The
second and the third states are nearly degenerate in mass. The fourth state
is dominated by 4-gluon configurations.  Future efforts will expand the
basis spaces by raising the regulators and begin to search for 
coherent phenomena such as those observed in scalar field
models mentioned above.
\section{Summary and Outlook} \label{sec 10}
While recent progress in the development and application of BLFQ has 
been encouraging, there is much work ahead to realize its potential.
At this intermediate stage, we observe that models with input from QCD show
many potentially fruitful research paths.  Applications of these BLFQ models
to mesons and baryons have revealed good agreement with results from
other approaches and with experiment for a multitude of observables. 
Applications to scattering of quarks in strong, time-dependent, external fields 
at the amplitude level also show good agreement with other models and with experiment.
Increasing the number of constituents in the Fock sectors and improving
understanding/implementation of renormalization are high on the priority list
for future efforts. The advance in supercomputer architecture and in linear
algebra algorithms will likely open additional horizons for the development and 
application of BLFQ to hadronic systems.
\begin{acknowledgements}
This work was supported by the Department of Energy under Grant Nos. DE-FG02-87ER40371 
and DESC000018223 (SciDAC-4/NUCLEI).
Computational resources were provided by NERSC, 
which is supported by the Office of Science of the U.S. DOE under Contract No. DE-AC02-05CH11231. Y. Li is supported in part by US DOE under Grant No. DE-FG02-04ER41302. X. Zhao is supported by new faculty startup funding by the Institute of Modern Physics, Chinese Academy of Sciences. We thank Robert Basili for assistance in creating Table I.
\end{acknowledgements}

\end{document}